\documentclass[12pt,preprint]{aastex}

\usepackage{morefloats}
\usepackage{color}

\shorttitle{Stellar Evolution and the Effects of Variable Composition on Habitable Zones}
\shortauthors{Truitt et al.}

\begin{document}

\title{A Catalog of Stellar Evolution Profiles and the Effects of Variable Composition on Habitable Systems}

\author{Amanda Truitt\altaffilmark{1}, Patrick A. Young\altaffilmark{1}, Alexander Spacek\altaffilmark{1}, Luke Probst\altaffilmark{1}, \& Jeremy Dietrich\altaffilmark{2}}
\altaffiltext{1}{School of Earth and Space Exploration, Arizona State University, Tempe, AZ 85287}
\altaffiltext{2}{Department of Astronomy, Harvard University, Cambridge, MA, 02183}

\newcommand{\Msol}{\mbox{$\rm{M_{\odot}\ }$}}
\newcommand{\msol}{\mbox{$\rm{M_{\odot}\ }$}}
\newcommand{\Rsol}{\mbox{$\rm{R_{\odot}\ }$}}
\newcommand{\rsol}{\mbox{$\rm{R_{\odot}\ }$}}
\newcommand{\sol}{\mbox{$\rm{_{\odot}\ }$}}
\newcommand{\Ni}{\mbox{$^{56}$Ni}}
  
\begin{abstract}    

We present stellar evolution models for 0.5 - 1.2 \Msol at scaled metallicities of 0.1 - 1.5 Z\sol and O/Fe values of 0.44 - 2.28 O/Fe\sol. The time dependent evolution of habitable zone boundaries are calculated for each stellar evolution track based on stellar mass, effective temperature, and luminosity parameterizations. The rate of change of stellar surface quantities and the surrounding habitable zone position are strong functions of all three quantities explored. The range of orbits that remain continuously habitable, or habitable for at least 2 Gyr, are provided. The results show that the detailed chemical characterization of exoplanet host stars and a consideration of their evolutionary history are necessary to assess the likelihood that a planet found in the instantaneous habitable zone has had sufficient time to develop a biosphere capable of producing detectable biosignatures. This model grid is designed for use by the astrobiology and exoplanet communities to efficiently characterize the time evolution of host stars and their habitable zones for planetary candidates of interest.

\end{abstract}


\section{INTRODUCTION}

One of the fundamental reasons for planet searches, and indeed one of the central motivations in the field of astronomy, is the eventual discovery of life on a planet outside our solar system. Though stellar evolution is not the only component to consider when working to understand planetary habitability, it is nonetheless one of the most important and physically well-understood factors. It plays a huge role in creating a ``habitable'' environment by providing energy to orbiting planets. Current astrophysical research into the habitability of exoplanets focuses mostly on the concept of the classical habitable zone (HZ), the range of distances from the star over which liquid water could exist on a planet's surface \citep{kwr93}. The location of the HZ is determined on the stellar side primarily by the host star's luminosity and secondarily by its spectral characteristics. These properties serve as boundary conditions for a planetary atmosphere calculation that predicts a planet's surface temperature, and therefore the possibility for stable liquid water. 

Over time, our understanding of the HZ has become more sophisticated as models of planetary atmospheres and their interaction with incoming stellar radiation have advanced. There are several components we must consider in order to accurately gauge the surface conditions of a planet, including: planetary atmospheric composition, cloud cover, and whether certain features (e.g. H{$_2$}O absorption bands) could be detected in atmospheric spectra \citep{vP11}; the physics of CO{$_2$} cloud condensation and evaporative water-loss; geophysical exchange with an atmosphere; planet mass, density, and surface temperature; and assumptions about greenhouse effects \citep{kwr93,sel07,kopp14}. For this work we use reasonably conservative estimates for the inner and outer boundaries of the HZ, based on HZ limit equations derived from a radiative-convective, 1-D atmospheric code that includes updated models of water absorption in planetary atmospheres, discussed in \citet{kopp14} (see \S 2.3).

It has become commonplace to announce the discovery of a planet when it is found in the HZ of its host star. A test of instantaneous habitability requires over-plotting a planetary orbit on the inner and outer boundaries of the HZ corresponding to observed temperature and luminosity of the star. This is not as straightforward as it would appear, since uncertainties in the stellar luminosity (often from distance uncertainties) can cause the HZ to move more than its entire width \citep{kane14}. Nonetheless, for stars with well-measured parameters we can place a planet within the HZ, and we can estimate the number of planets that should be in the HZ in a statistical sense. The {\it Kepler} mission was launched in 2009 with an explicit goal of finding an Earth-sized planet orbiting within the HZ, and was very successful in discovering these small terrestrial objects \citep{borucki11,char11,fress12}. Subsequent data analysis has predicted that one in five Sun-like stars \citep{catshao11,pet13,gaidos13} and approximately half of all M-dwarf stars \citep{borucki10,bat13} are likely to host an Earth-sized planet in the HZ. 

For astronomers hoping to find planets with detectable biosignatures, this introduces the novel problem of having {\it too many} candidates. Directly detecting and obtaining spectra for an Earth-like planet around a Sun-like star is typically difficult and would likely require a large-scale space mission. Such a mission may be able to observe no more than about a hundred stars over its lifetime (e.g. \citet{turn12}). As a first step, it would be useful to narrow-down the planetary candidates and associated host stars on the basis of whether a planet is $currently$ habitable, which we can estimate directly from the measured properties of the star; however, the current habitability of a planet is not the only important consideration for finding Earth-like worlds. If the ultimate goal is the discovery of life, the evolution of stars {\it and} the associated HZ must be taken into account as well. 

The likelihood for detection of life on a habitable planet ultimately stems from a probability distribution constructed of multiple terms falling into three broad classes: the ability of a planet to sustain a biosphere, the likelihood that biosignatures can persist at detectable levels on the planetary surface, and the technique being used to observe the planet. We can begin defining individual terms for each of these categories, which will be expanded upon as our understanding progresses. As examples of the latter classes, some inhabited planets will probably be unidentifiable until we are able to examine them from up close. An obvious example is Europa in our own solar system, which may support a biota under its icy crust. The limited communication of the potentially life-sustaining region with the surface gives Europa a low potential for detectability because biosignatures will escape at low levels at best and will be rapidly destroyed by Jupiter's radiation environment.

More subtle effects may come into play that could even prevent detection of surface biospheres on terrestrial planets. As examples, planets with a highly reducing bulk chemistry could conceivably fix free oxygen faster than even a substantial population of photosynthesizing organisms could produce it, which could deprive us of an opportunity for detection of life. Free oxygen is often suggested as one of the more ``useful'' or ``indicative'' biosignatures, although recently that idea has been called into question; it has become more apparent that oxygen present in a planetary atmosphere may not necessarily indicate any sort of biogenic process \citep{domgold14,wordp14,luger15}. Such a planet might, on the other hand, be able to sustain high levels of biogenic methane that could not be attributed to geological processes such as serpentinization. This planet would have a higher likelihood of detection for missions that are able to observe methane as well as oxygen, which is why a thorough characterization of the entire planetary system is important.

One of the first things we should consider in attempting to understand a whether a planet may support life, and the likelihood of detection, must be the characteristics of the host star and the location of the surrounding HZ. We have the soundest basis for developing detection strategies for Earth-like life, and such life is clearly favored by Earth-like conditions. Such conditions are, by definition, most likely to be found in the HZ. However, the instantaneous location of the planet and HZ are too limited to assess this aspect of the planet's habitability. The location of the HZ is a function of time; we know that as stars age, the habitable region will move away from the star due to the gradual increase in stellar luminosity ($L$). The effective temperature ($T_{eff}$) will also evolve, changing the spectrum impinging upon the planet's atmosphere. Therefore, any orbiting planets detected around a particular star may only have spent a relatively short time in the HZ.  It is easy to assume that a planet around a sedate, five billion year old star will be as rich and diverse as Earth. Time dependent models show, however, that more than half of the orbits that are in the HZ at some point during the evolution are actually only habitable during the latter part of the star's life. Our hypothetical five billion year old planet will not be a very fecund place if it just entered the HZ 100 Myr ago. It is estimated that life only produced a detectable change in Earth's atmospheric chemistry 1 -- 2 Gyr after Earth's formation \citep{kast93,bro99,kopp05,anbar07,crowe13} even though Earth was in the Sun's HZ from very early times.

Identifying these kinds of systems from readily measurable stellar and orbital parameters would rapidly reduce the number of habitable planetary candidates. An essential parameter that affects stellar and HZ evolution is the {\it detailed} chemical composition of a star. The rate of stellar evolution and the change in $T_{eff}$ and $L$ are dependent on the abundances of individual elements, especially oxygen and iron \citep{ylp12}. The sensitivity of stellar evolution to composition arises from two effects: the equation of state (EOS) and the radiation opacity \citep{rsi96,ir96}. The changes in the EOS are relatively minor, but rearranging the proportions of different species at a constant [Fe/H] can result in significant opacity changes of tens of percent \citep{ir96}.  Opacity changes the rate of leakage of radiation, and increased radiation pressure in the stellar envelope drives expansion, resulting in larger radii and lower effective temperatures. A slower rate of energy loss also requires a slower rate of nuclear burning to maintain hydrostatic equilibrium. So we expect stars with enhanced abundance ratios [X/Fe] to be cooler, less luminous, and longer-lived relative to stars with the same [Fe/H].

Measurements of [Fe/H] alone are insufficient to predict the stellar and HZ evolution \citep{ylp12}. In practice, it is usually only the iron abundance that is measured for many stars, and the rest of the elements are assumed to scale in the same proportions present in the Sun; this means models of stellar evolution for stars of different metallicity are generally created under this assumption. This is the nearly universal practice in stellar modeling, though the abundance ratios in real stars vary substantially. Except for a uniform enhancement of the $\alpha$ elements (O, Ne, Mg, Si, S, Ar, Ca, Ti) in stars of very low metallicities, variations in abundance ratios at a given [Fe/H] are neglected, despite the fact that stellar evolution is sensitive to specific composition.    

For solar mass stars near solar [Fe/H], in all cases, enhanced abundance ratios cause larger changes than depleted ratios. C, Na, and Mg have small but noticeable effects. C has a high abundance but relatively few electron transitions and a low ionization potential. Mg and Na have lower abundances but higher opacity per gram than C, resulting in a similar degree of shift in the tracks. Si has less impact due to its smaller range of variation, and Al, Ca, and Ti have very small effects due to their small abundances; these can be neglected for the purpose of habitability. The largest changes arise from variation in oxygen \citep{ylp12}. Both $L$ and $T_{eff}$ of the enriched compositions are systematically lower at a given age, but stellar evolution models show the most profound effect is on the pace of the evolution. For example, for a 1 M\sol star with solar [Fe/H] the main sequence (MS) turnoff for a model representing the low end of the distribution of oxygen abundance in nearby stars occurs at an age of $\sim$ 9 Gyr. Solar composition has a turn-off age of $\sim$ 10 Gyr, and an oxygen-rich model turns off at $\sim$ 11 Gyr (and at lower $L$ and $T_{eff}$).

In this paper, we examine how stellar mass, metallicity, and individual elemental abundance ratios influence the long term habitability of planetary systems. We present a grid of stellar evolution models for Sun-like stars, with masses of 0.5 - 1.2 M\sol, metallicities of Z = 0.1 - 1.5 Z\sol, and a spread in oxygen values ranging from O/Fe = 0.44 - 2.28 O/Fe\sol. When calculating the time-dependent HZ inner and outer radii for each of our stellar evolutionary tracks, we have chosen to use a widely cited and well-known set of prescriptions \citep{kopp14}, though the main utility of our catalog is that our models are easily transferrable for use with any particular set of HZ limit equations; this includes scenarios where a so-called ``cold start'' is not viable (i.e. a frozen planet enters the HZ late in its evolution, yet does not have the capability to thaw for the sake of habitability). \citet{kopp14} work under the assumption that a cold start is in fact plausible. Following \citet{kwr93} we present a scenario where the outer limit of the HZ at the zero age main sequence (ZAMS) is a hard limit that does not vary with the star's evolution so that these two scenarios can be compared.

We examine F, G, and K-type stars in particular because they are the closest in physical properties to our own Sun, which is a good starting point of comparison if we want to look for potential host stars of habitable Earth-like planets. M-stars present other issues for habitability, including high levels of stellar activity with high energy particle and x-ray fluxes, close-in HZs with potential for tidal locking, spin-orbit resonances and tidal heating, and dynamically packed inner systems. We therefore defer calculations of M-stars, which require more comprehensive treatments of complex molecular opacities, to a separate paper.

Ultimately, we would like to learn what kind of star could potentially host a planet that has remained in a continuously habitable zone (CHZ) for at least 2 Gyr. This is approximately the amount of time it took for life on Earth to change the atmospheric composition sufficiently that there would be a detectable biosignature if viewed from another star system, utilizing missions recommended in the most recent Decadal Review of Astronomy and Astrophysics (e.g. transmission spectroscopy with JWST or direct detection with a coronagraph, starshade, or interferometer). We define a CHZ for the star's entire MS lifetime, as well as the range of distances that would be continuously habitable for at least 2 Gyr at {\it some period} in the star's evolution. We describe our choice of parameter space, stellar evolution code, and assumptions for calculating the HZ in \S 2, our interpretation of the results of the models in \S 3, and our conclusions in \S 4.

\section{METHODS}
\subsection{Parameter Space \label{sec.ps}}

In this work we present a grid of stellar models suitable for the prediction of HZ locations. The most important variable for stellar evolution is, of course, mass. Second is total metallicity, Z. Third is the oxygen abundance ratio, O/Fe. The ratios C/Fe and Mg/Fe also produce a small effect in stellar evolution and have substantially variable abundance ratios \citep{neves09,mish08,take07,young14}. These will be explored in a future paper. In this work, ratios without brackets (e.g. O/Fe) indicate the linear absolute abundance ratio in terms of mass fraction, while a bracketed ratio denotes the log of the atom number relative to the solar abundance value for that same element. The latter is the conventional [O/Fe] given by
\begin{equation}
log_{10} \frac{(O/Fe)}{(O/Fe)\sol} = [O/Fe]
\end{equation}
This work primarily quotes linear ratios relative to solar (i.e. O/Fe = 2.28 O/Fe\sol) since the range of abundance ratios is small enough to not require logarithmic notation. We use mass fraction as this is the conventional usage for stellar evolution calculations.

Here we consider the major contributors, mass, metallicity, and oxygen abundance. Variations in Z alone are made with a fixed abundance pattern that is uniformly scaled. The spread in oxygen values we use reflects actual variations in oxygen abundances that have been directly observed in nearby stars \citep{ram07,bond06,bond08,gonz10,hink14}. The values of O/Fe\sol are taken from \citet{young14}, which analyzes the intrinsic spread (not accounted for by observational error) in elemental abundance ratios [X/Fe] for 5 large surveys. We use the [O/Fe] from the median survey, which is consistent with the values used for the calculations in \citet{ylp12}. A solar composition from \citet{lod10} was adjusted to the mean abundance ratios of the median sample for the elements observed, with all other elements being maintained at solar values for standard composition. The adjusted values are listed in Table~\ref{tab1}. Changes in O/Fe\sol at each metallicity are made by changing the absolute abundance of O while holding all other metal abundances constant. The abundances of H and He are adjusted in compensation to ensure the sum of mass fractions = 1.

Determinations of O abundance are notoriously sensitive to non-LTE effects \citep{gas07} and line-blending, which may introduce random unphysical errors in the O abundance measurement that are not accounted for in the quoted observational errors. The surveys considered in \citet{young14} were chosen in part for using the most reliable of the optical O lines, but we also examine a smaller range of O variation. Instead of choosing a random value, we use the $u_{intrinsic}$ \citep{young14} for a sample of 40 ``solar twins'' from \citet{ram09}. Qualifying as solar twins requires observable parameters quantitatively close to the Sun. In order for a star to match the Sun closely in physical observables, it is necessary for it to have a composition close to solar. Using $u_{intrinsic}$ for the solar twin sample thus provides a conservative estimate of [O/Fe] variation.

The initial grid encompasses solar-type stars on the MS, covering a mass range from 0.5 - 1.2 M\sol at increments of 0.1 M\sol, corresponding to spectral types M0 - F0 at Z = Z\sol. Models are calculated for metallicities of 0.1 - 1.5 Z\sol at increments of 0.1 Z\sol. Two models are calculated with values of O/Fe\sol at each Z value (end members 2.28 and 0.44 O/Fe\sol). The other two O cases (0.67 and 1.48 O/Fe\sol) were done only at solar Z value. The grid is complete for the MS until hydrogen exhaustion in the core for all cases. Post-MS evolution for short lived stars ($>$ 1 M\sol), cool stars (e.g. M-dwarf stars), and contributions from minor elemental constituents (such as C and Mg) will be explored in a future publication.

\subsection{TYCHO}

The stars included in our catalog were simulated using the stellar evolution code TYCHO \citep{yoar05}. TYCHO is a 1-D stellar evolution code with a hydrodynamic formulation of the stellar evolution equations.  It uses OPAL opacities \citep{ir96,alfer94,rn02}, a combined OPAL and Timmes equation of state (HELMHOLTZ) \citep{ta99,rn02}, gravitational settling (diffusion) \citep{tbl94}, general relativistic gravity, time lapse, curvature, automatic rezoning, and an adaptable nuclear reaction network with a sparse solver. A 177 element network terminating at $^{74}$Ge is used throughout the evolution. The network uses the latest REACLIB rates \citep{rat00,ang99,iliadis01,wie06}, weak rates from \citet{lmp00}, and screening from \citet{grab73}. Neutrino cooling from plasma processes and the Urca process is included. Mass loss is included but is trivial for this mass range. (Heightened early mass loss seen in some young stars \citep{wood05} is not included.) It incorporates a description of turbulent convection \citep{ma07,amy09,amy10,am11}, which is based on 3-D, well-resolved simulations of convection between stable layers,  which were analyzed in detail using a Reynolds decomposition into average and fluctuating quantities. It has no free convective parameters to adjust, unlike mixing-length theory.

TYCHO outputs information on stellar surface quantities (including $L$ and $T_{eff}$) for each time-step of a star's evolution, which we then use to calculate the inner and outer radii of the HZ as a function of the star's age. The initial composition of the stellar models were adjusted as described in \S 2.1, and new OPAL opacity tables were generated for the specific abundances for each O/Fe value to match the composition of the stellar model. TYCHO begins calculation with a fully convective model on the Hayashi track. We limit our discussion of the HZ to the span between the Zero Age Main Sequence (ZAMS) and the Terminal Age Main Sequence (TAMS). For this purpose, the ZAMS is defined as the luminosity minimum coinciding with the beginning of complete hydrogen burning in the core. The TAMS is defined as the time of hydrogen exhaustion in the core ($X_H < 1.0\times 10^{-6}$). The complete tracks are publicly available.\footnote{http://bahamut.sese.asu.edu/$\sim$payoung/AST\_522/Evolutionary\_Tracks\_Database.html}

\subsection{Calculating Habitable Zone Extents}

TYCHO evolutionary tracks are used to estimate the extent of the HZ at each point in the stellar evolution. For these estimates we follow the prescriptions of \citet{kopp13,kopp14}, which follow from \citet{sel07} and \citet{kwr93}. These prescriptions parameterize the orbital radii of the HZ as a function of $L$ and $T_{eff}$ facilitating translation from stellar evolution tracks to HZ distance estimations. 

\citet{kopp14} use radiative-convective planetary atmosphere models with input synthetic stellar spectra produced by the PHOENIX code \citep{ah95}. The predicted distance from a star for both the inner and outer edges of the HZ is parameterized as a function of stellar temperature and luminosity, for several kinds of planetary atmospheres. The calculations do not include other sources of heating such as tides (small for these stars) or effects such as eccentric orbits. The distance {\it d} of the HZ boundary is defined by
\begin{equation}
d = \left(\frac{L/L_{\odot}}{S_{eff}}\right)^{1/2}
\end{equation}

where $S_{eff}$ is an effective flux received by the planet based on the radiative transfer calculations for different initial spectra corresponding to different $T_{eff}$ passing through the planetary atmosphere. 
\begin{equation}
S_{eff} = S_{eff\odot} + aT_{*} + bT_{*}^2 + cT_{*}^3 + dT_{*}^4
\end{equation}
Where {T$_{*}$} = {T$_{eff}$} - 5780 K, $S_{eff\odot}$ is the effective flux for Earth from the Sun at the solar $T_{eff}$,  and $a, b, c,$ and $d$ are coefficients to a polynomial fit. Each distinct planetary atmosphere model generates a unique set of coefficients. 

Five cases are described in \citet{kopp13}: (1) Recent Venus, (2) Runaway Greenhouse, (3) Moist Greenhouse, (4) Maximum Greenhouse, and (5) Early Mars. However, \citet{kopp14} replaces the Moist Greenhouse case entirely with the Runaway Greenhouse case, since the differences between the two are minimal. We use the updated values for the coefficients (a, b, c, d) and {S$_{eff\odot}$} from \citet{kopp14}.

The four remaining cases represent a significant range of HZ approximations when we consider both the stellar flux incident on the planet due to the planet's distance from the host star, along with several different atmospheric properties that may exist on an Earth-like planet. Though the amount of solar radiation the planet receives is the largest factor in determining HZ distances, the wavelength-dependent radiative transfer and radiative losses from atmospheres with different levels of greenhouse gases, water, and clouds must also be considered. HZ boundaries are undoubtedly strongly influenced by the presence of clouds in a planet atmosphere. We would expect H{$_2$}O clouds to move the inner HZ boundary inward \citep{kast88,sel07,yang13} because their contribution to a planet's albedo generally outweighs the contribution to the greenhouse effect. Conversely, CO{$_2$} ice clouds are expected to cause warming in a dense CO{$_2$} atmosphere because they reflect thermal radiation back to the planet's surface more efficiently than they reflect incoming radiation back to space \citep{fp97}.

The Runaway Greenhouse (RGH) and Maximum Greenhouse (MaxGH) cases represent the conservative estimates for the inner and outer radii of the HZ, respectively. The RGH limit is the distance at which a planet's oceans would evaporate entirely. The inner edge of the HZ is determined by the level of water saturation in the planet's atmosphere (a warm environment causes evaporation of H{$_2$}O from the planet's surface) and the subsequent rapid loss of hydrogen to space due to heating effects from the proximity of the host star. Conversely, planets that orbit near the outer HZ boundary could develop dense, CO{$_2$}-rich atmospheres through outgassing. CO{$_2$} begins to condense out of the atmosphere at a certain distance away from the parent star (due to colder temperatures) which reduces the overall greenhouse effect. CO{$_2$} is also an effective Rayleigh scatterer (2.5 times better than air), and so a dense CO{$_2$} atmosphere is expected to have a high albedo which would further offset the greenhouse effect \citep{kast91}. Thus, the conservative outer HZ boundary (MaxGH) is the location where Rayleigh scattering by CO{$_2$} begins to outweigh the greenhouse effect.

Additionally, the Recent Venus (RV) case predicts an inner HZ edge much closer to the star, while the Early Mars (EM) case predicts a more distant outer boundary. These we will refer to as the optimistic cases. The RV case limit from radar observations of Venus by the {\it Magellan} spacecraft that suggest liquid water has been absent from the surface of Venus for at least 1 Gyr \citep{sohead91}, when the Sun was about 90\% as luminous as it is today. This gives a more optimistic empirical estimate for the inner HZ boundary. Likewise, the (optimistic) Early Mars (EM) outer boundary has been estimated based on the observation that early Mars was warm enough for liquid water to flow on its surface \citep{pol87,bib06}. Although the issue of a warmer early Mars has raised some debate \citep{seg02,seg08}, this gives a good estimate of a more distant outer boundary at which a planet could potentially remain habitable.

The TYCHO evolutionary tracks are used as input to CHAD (Calculating HAbitable Distances), which is a code we developed to calculate the inner and outer HZ boundaries for each of the HZ limits presented in \citet{kopp14}. CHAD is easily upgradable to incorporate improved HZ predictions as they become available in the future. In this paper we will focus on the conservative estimates, but all of the calculations will be made publicly available in machine-readable format along with the evolutionary tracks.

\section{RESULTS}

From a stellar point of view, three main factors influence the time evolution of the HZ. These are $L$ and $T_{eff}$, their rate of change, and the stellar MS lifetime. We confirm that mass and total metallicity influence these factors considerably. Following on \citet{ylp12}, the ratio of O/Fe is also significant over the entire range of M and Z. When we consider the evolution of the HZ, it is clear that the instantaneous HZ calculated from observed stellar properties is often not a good indicator of a high likelihood for detection of biosignatures, except for very low mass stars. Detectability ultimately depends on assumptions made about the timescales involved in evolution of life and biogeochemical evolution of biosignatures.

\subsection{Stellar Properties and Main Sequence Lifetimes}

\begin{figure}
\centering
\includegraphics[width=1\textwidth]{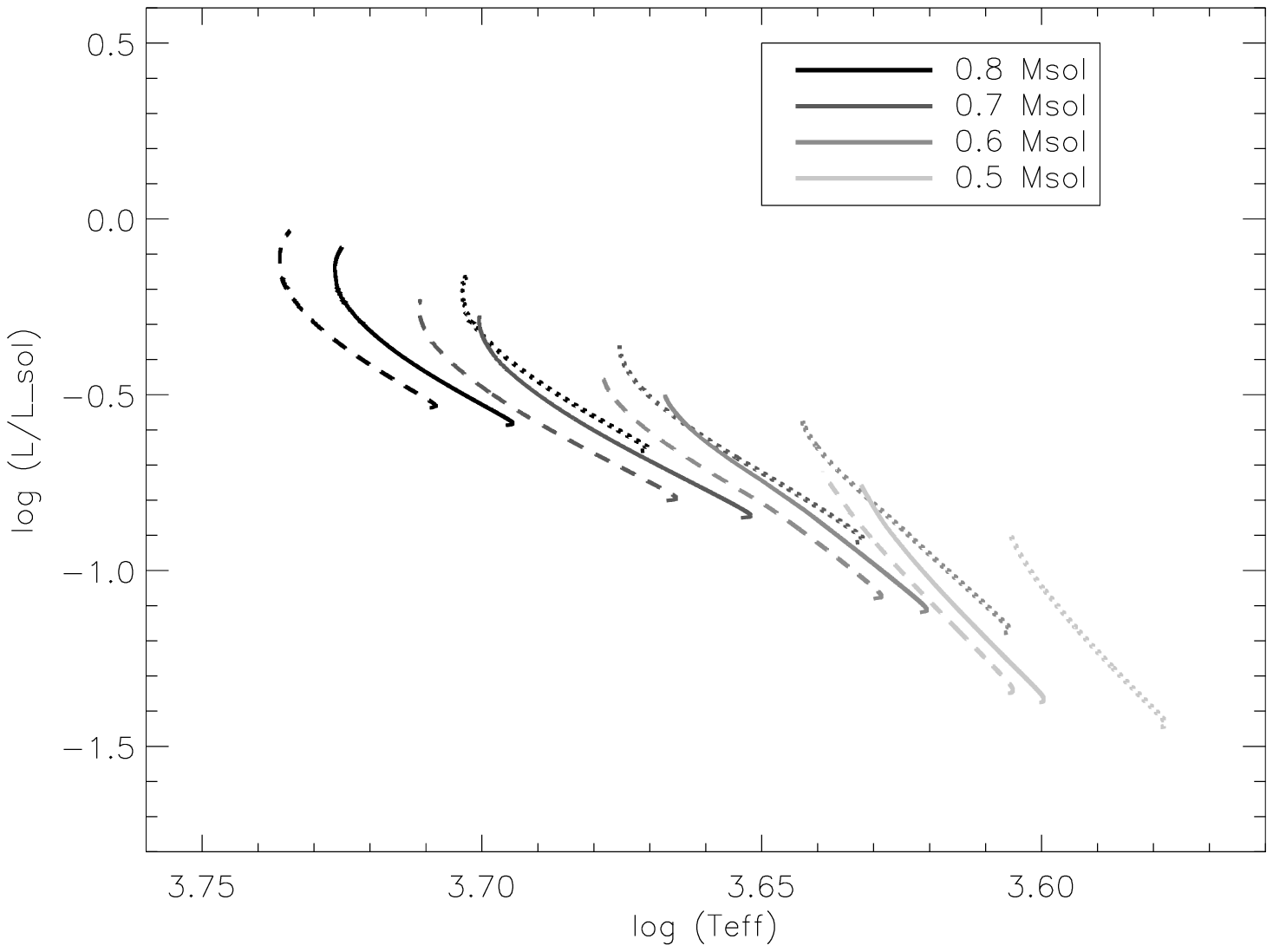}
\caption{HRD, Evolutionary tracks from ZAMS to TAMS for 4 masses, with O/Fe = 0.44 O/Fe\sol (dashed), 1.0 O/Fe\sol (solid), and 2.28 O/Fe\sol (dotted) all at Z = Z\sol. The rightward-most dotted line is for the 0.5 M\sol star with enriched oxygen, while the leftward-most dashed line is for the 0.8 M\sol star with depleted oxygen.\label{hrd1}}
\end{figure}

\begin{figure}
\centering
\includegraphics[width=1\textwidth]{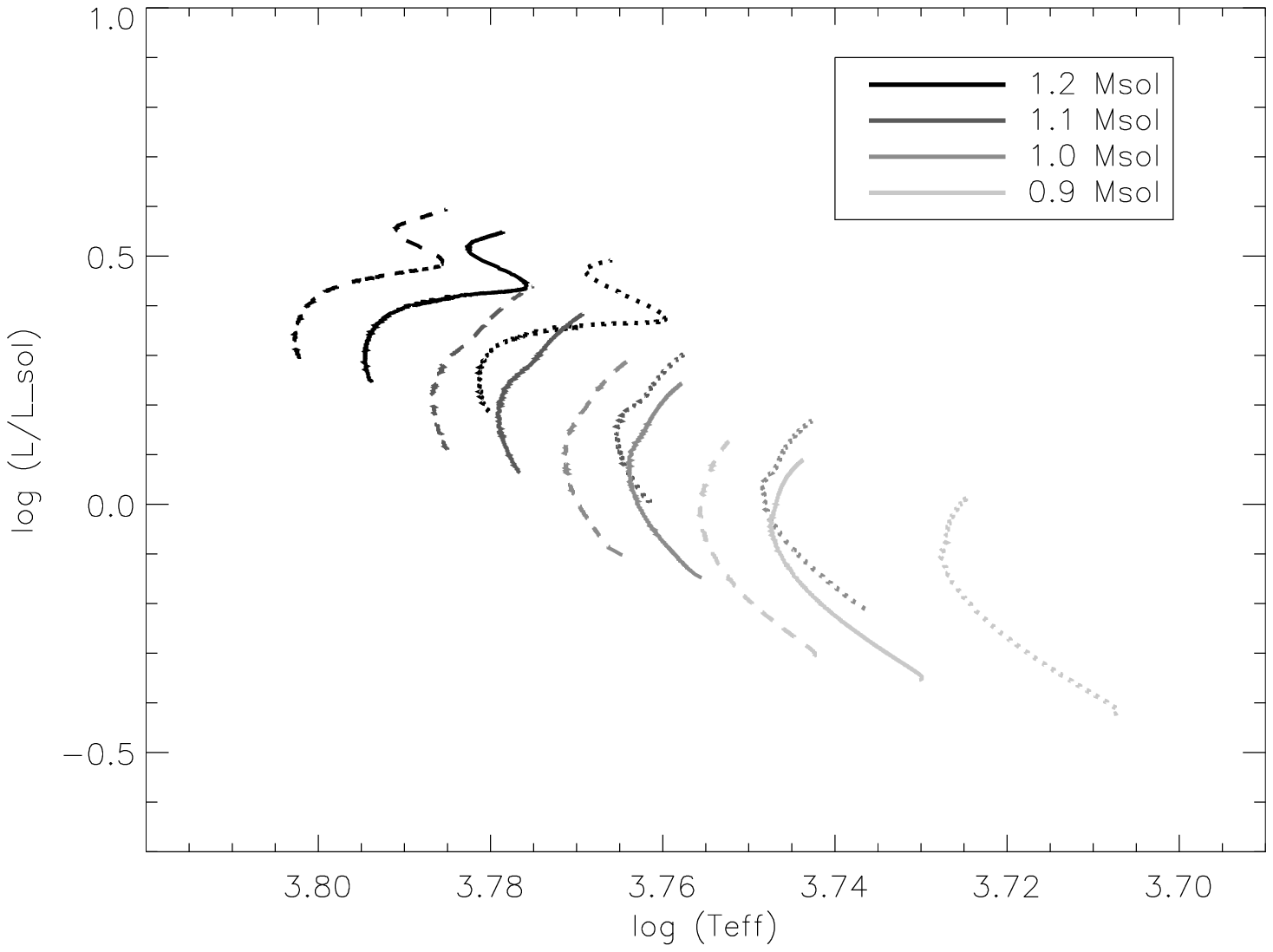}
\caption{HRD, Evolutionary tracks from ZAMS to TAMS for 4 masses, with O/Fe = 0.44 O/Fe\sol (dashed), 1.0 O/Fe\sol (solid), and 2.28 O/Fe\sol (dotted) all at Z = Z\sol. The rightward-most dotted line is for the 0.9 M\sol star with enriched oxygen, while the leftward-most dashed line is for the 1.2 M\sol star with depleted oxygen.\label{hrd2}}
\end{figure}

TYCHO stellar evolution tracks are used to determine the stellar parameters of interest at each time-step in a star's evolution. The evolution was calculated from the pre-MS to H exhaustion. The pre-MS evolution is rapid, with a similar timescale to planet formation, and can be neglected. This provides us with a MS lifetime and a rate of change for the stellar properties. The Hertzsprung-Russell diagrams (HRD) shown in Figures~\ref{hrd1} and~\ref{hrd2} demonstrate the effect of variations in oxygen abundance ratios for each mass in our data set, from 0.5 M\sol to 0.8 M\sol in Figure~\ref{hrd1}, and from 0.9 M\sol to 1.2 M\sol in Figure~\ref{hrd2}. For both figures, the solid lines represent solar-value for the O/Fe ratio, whereas the dashed lines and dotted lines correspond to depleted and enriched oxygen cases (0.44 O/Fe\sol and 2.28 O/Fe\sol), respectively. Significant changes are seen when the O abundance ratio varies, even at constant Z. Trends with O/Fe parallel those with the total metallicity; low O/Fe stars are bluer, more luminous, and shorter-lived. Higher O/Fe stars are cooler, less luminous, and longer lived. Thus, the specific O abundance in stars (not just the overall metallicity) plays a significant role in stellar evolution and, consequently, planetary habitability.

\begin{figure}
\centering
\includegraphics[width=1\textwidth]{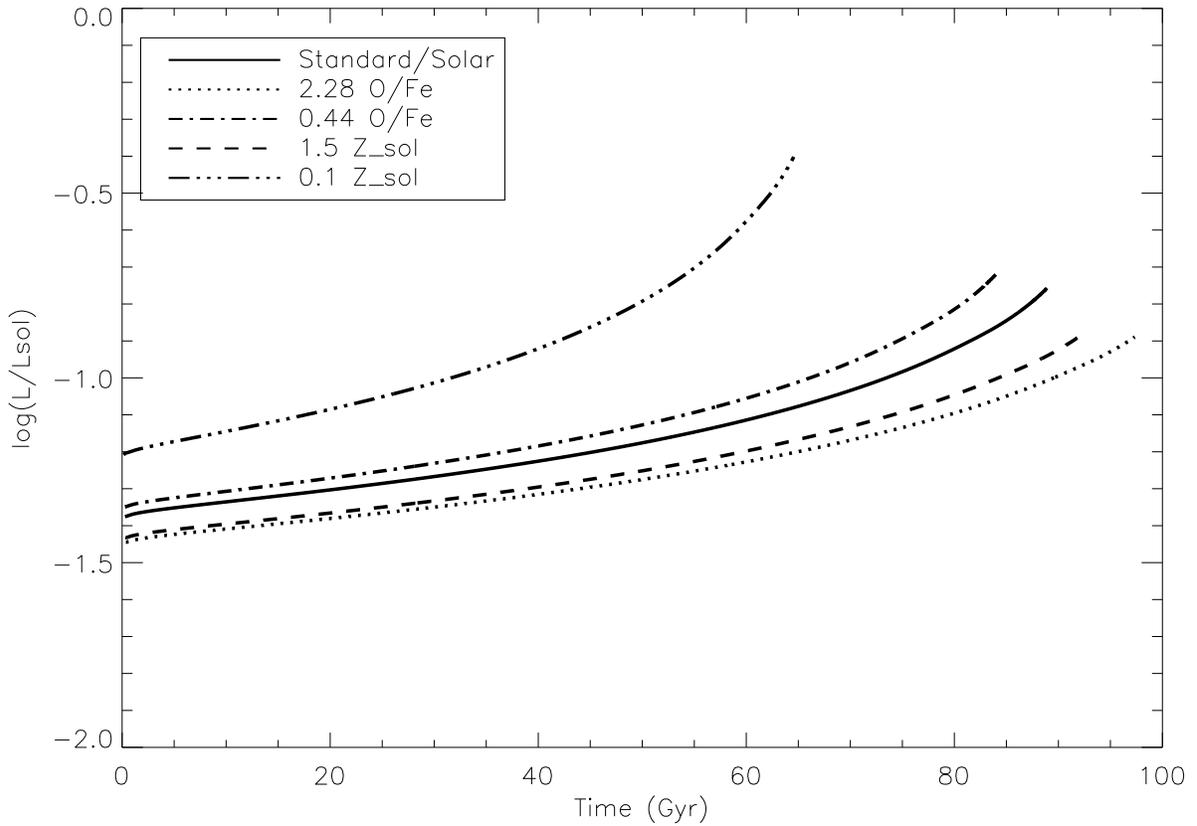}
\caption{Log($L$/$L$\sol) vs. time (Gyr) for a 0.5 M\sol star for five different compositions. The total MS lifetime varies from 65 Gyr to nearly 100 Gyr. The shortest lifetime corresponds to a star with total metallicity Z = 0.1 Z\sol. The longest lifetime corresponds not to the highest scaled Z model, but rather to the model with O/Fe = 2.28 O/Fe\sol (at Z = Z\sol).\label{5mass}}
\end{figure}

\begin{figure}
\centering
\includegraphics[width=1\textwidth]{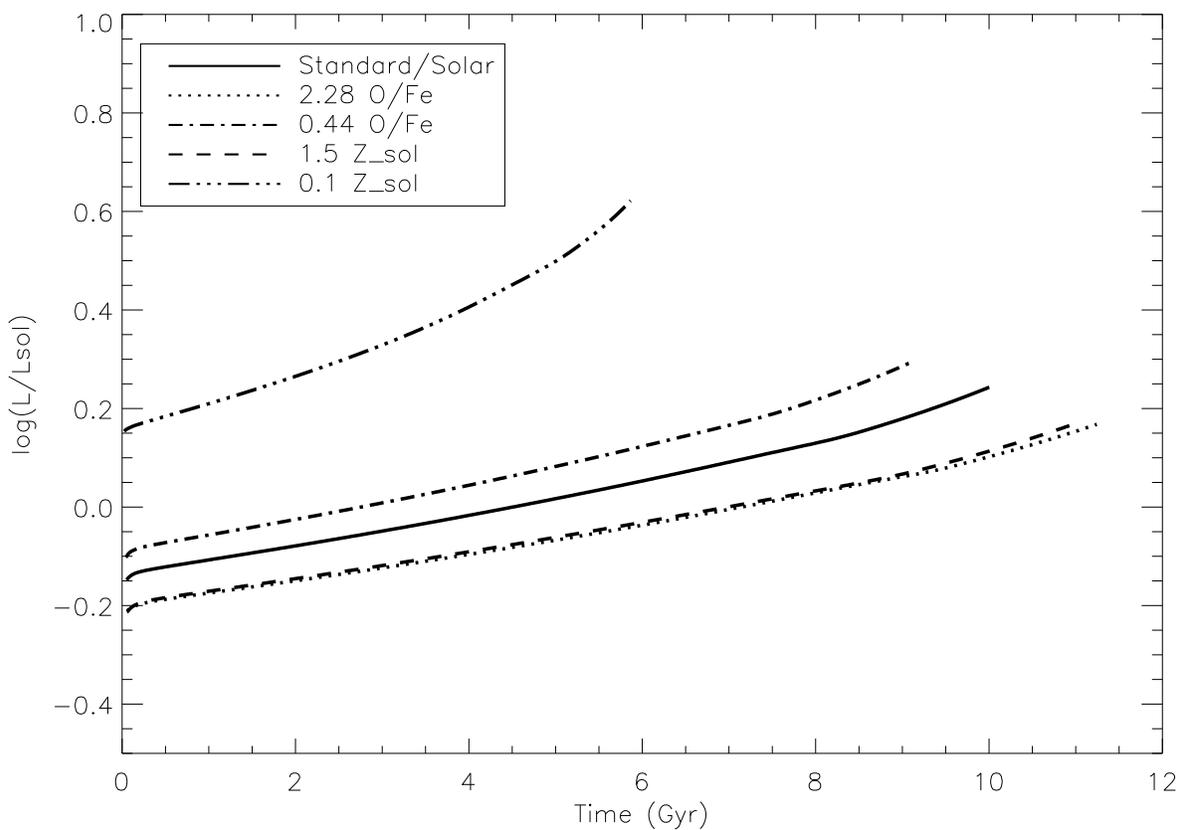}
\caption{Log($L$/$L$\sol) vs. time (Gyr) for a 1 M\sol star for five compositions. We see now that the total MS lifetime has a much smaller variation, from $\sim$ 6 - 11.5 Gyr. Again, the shortest lifetime corresponds to a star with metallicity Z = 0.1 Z\sol, and the longest lifetime corresponds to the model with O/Fe = 2.28 O/Fe\sol (at Z = Z\sol).\label{1mass}}
\end{figure}

\begin{figure}
\centering
\includegraphics[width=1\textwidth]{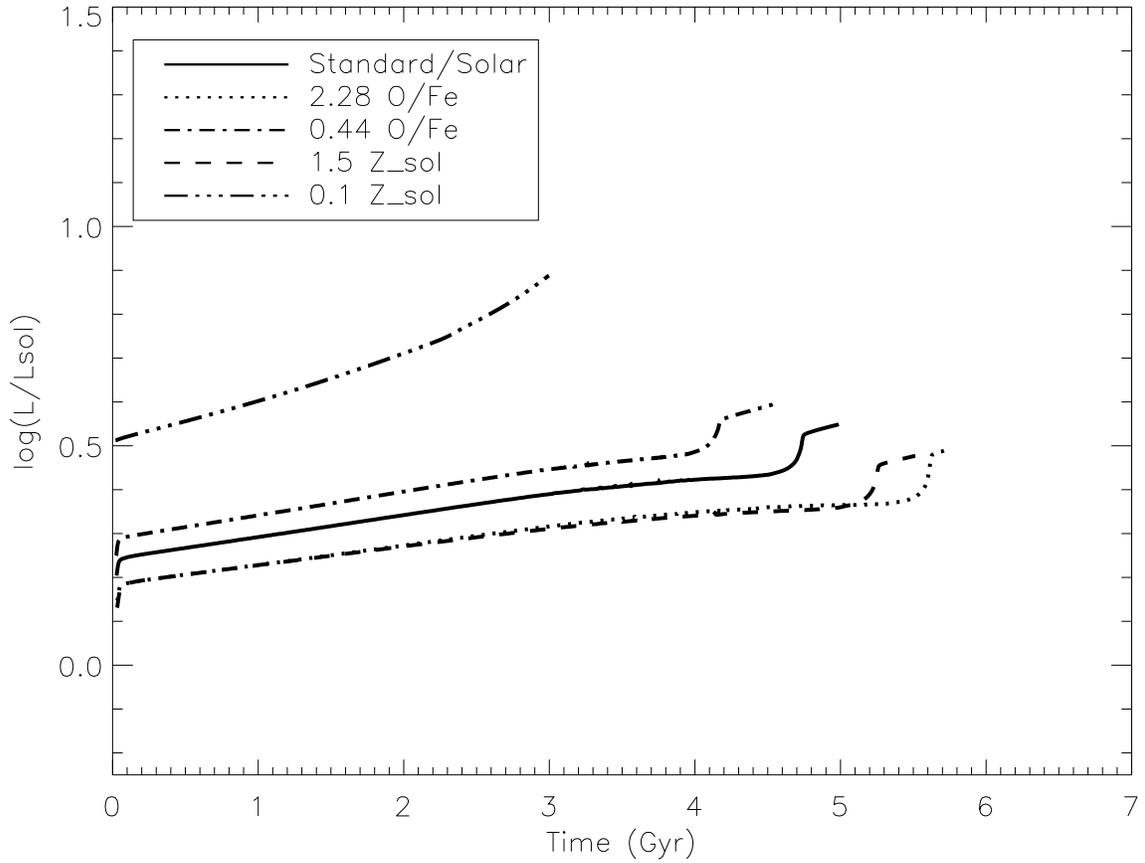}
\caption{Log($L$/$L$\sol) vs. time (Gyr) for a 1.2 M\sol star for five compositions. Again, the shortest lifetime corresponds to a star with metallicity Z = 0.1 Z\sol, and the longest lifetime corresponds to the model with O/Fe = 2.28 O/Fe\sol (at Z = Z\sol).\label{2mass}}
\end{figure}

Figure~\ref{5mass} shows log($L$/$L$\sol) vs. time (Gyr) for the compositional end member cases in our library for a 0.5 M\sol star. Models for 0.1 Z\sol, 1.5 Z\sol, Z\sol, 0.44 O/Fe\sol and 2.28 O/Fe\sol (each at Z = Z\sol) are shown. Figures~\ref{1mass} and \ref{2mass} show the same for 1 M\sol and 1.2 M\sol stars. For all masses we see a significant variation in the MS lifetime, with the lowest metallicity star living just two thirds as long as the highest metallicity model for the 0.5 M\sol star, and about half as long as the highest metallicity model for the 1 M\sol star. Of particular importance is the change wrought by varying O/Fe. In this case, an otherwise solar composition star with O/Fe = 2.28 O/Fe\sol actually lives longer than a Z = 1.5 Z\sol star. This illustrates the danger in only measuring [Fe/H] and assuming that a scaled solar metallicity will accurately predict a star's evolution. For the 0.5 M\sol star, the high-end cases for O and Z are quite similar in terms of $L$, however the enriched O case sees a longer MS lifetime by about 6 Gyr. As expected, we see that stars with higher Z (or O/Fe) are less luminous, but considerably longer-lived. Table~\ref{tab2} shows MS lifetime estimates for the range encompassed by our catalog, acquired using this method.

A more subtle effect to consider is the rate of change of $T_{eff}$ and $L$. Low opacity models undergo a larger change in luminosity than do the higher opacity models at the same mass, over a shorter total lifetime. Thus, $dL/dt$ is greater for low metallicity (or O/Fe) models, especially during the second half of a star's evolution on the MS. The radial movement of the HZ boundaries is concomitantly faster. At 0.5 M\sol, a 1.5 Z\sol model increases in $L$/$L$\sol by 0.025 dex, while a 0.1 Z\sol model is brighter by 0.05 dex at the end of the MS. This translates to the luminosity of the star changing by a factor of 5.4 at 0.1 Z\sol compared to only 2.9 at 1.5 Z\sol. The absolute change in luminosity is even more sensitive to variations in composition at higher masses, though the percentage change is smaller. Tables~\ref{tab3} and \ref{tab4} show the total change in $L$ in terms of $L$/$L_{ZAMS}$, and the total change of $T_{eff}$ in units of Kelvin (respectively) from ZAMS to TAMS.

Finally, because of how the slopes of $dL/dt$ (and to a lesser extent $dT/dt$) change over time, the range of orbits in the HZ at different points in the MS evolution can change substantially. When considering the potential for detectability, we wish to avoid planets that have only recently entered their HZ. Table~\ref{tab5} shows the fraction (listed as percentages) of orbital radii which enter the HZ $after$ the midpoint of the MS. It is clear from these results that a third to a half of orbits that are in the HZ only become habitable in the second half of the star's life. The effect is more pronounced at higher mass and composition.

\subsection{Location of the Habitable Zone}

\begin{figure}
\includegraphics[width=1\textwidth]{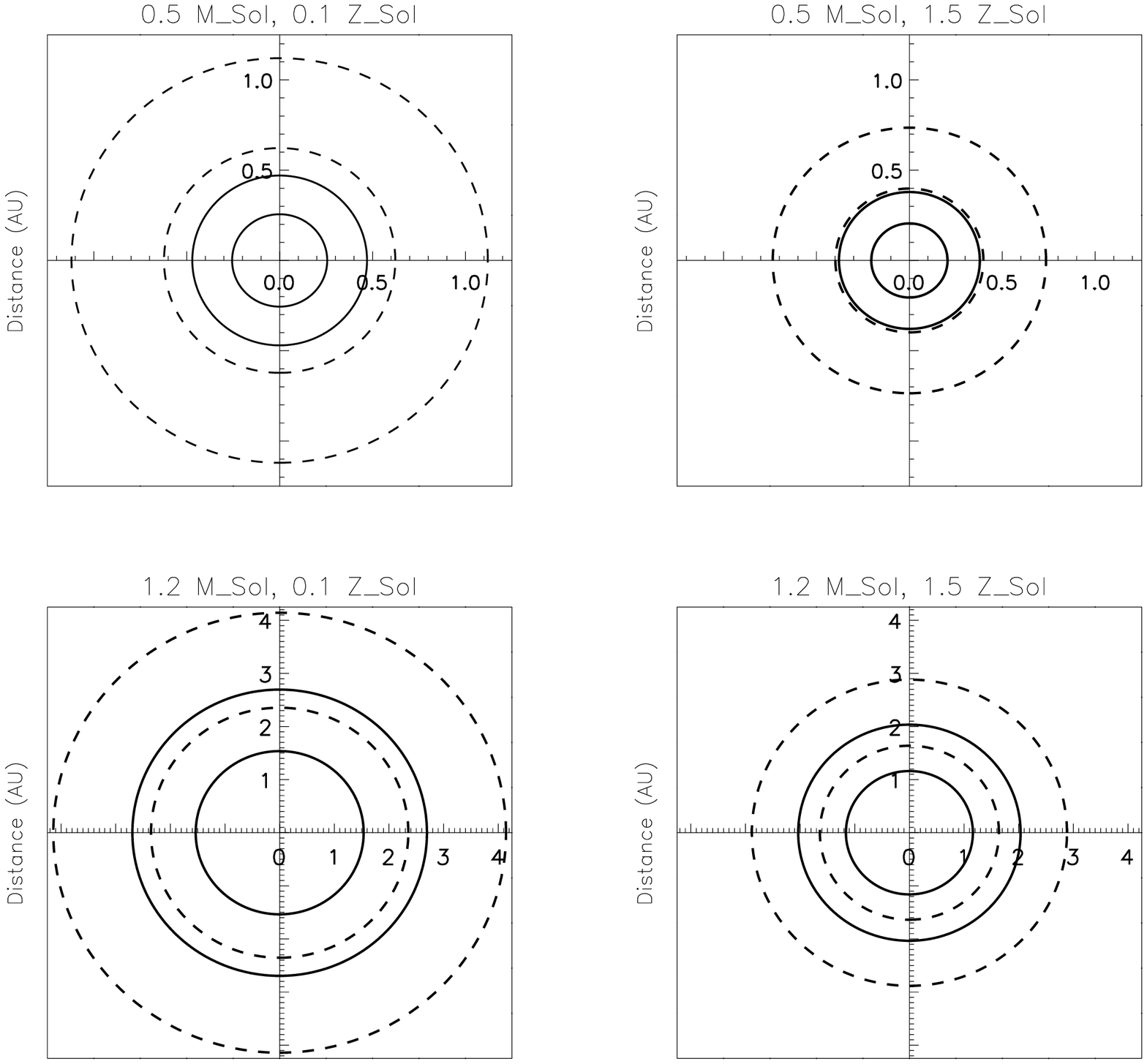}
\caption{Habitable Zone Ranges: 0.5 M\sol (top), 1.2 M\sol (bottom), for 0.1 Z\sol (left), 1.5 Z\sol (right). HZ shown at ZAMS (solid) and TAMS (dashed). Inner boundaries represent Runaway Greenhouse and outer boundaries represent Maximum Greenhouse.\label{hzpolar1}}    
\end{figure}

\begin{figure}
\includegraphics[width=1\textwidth]{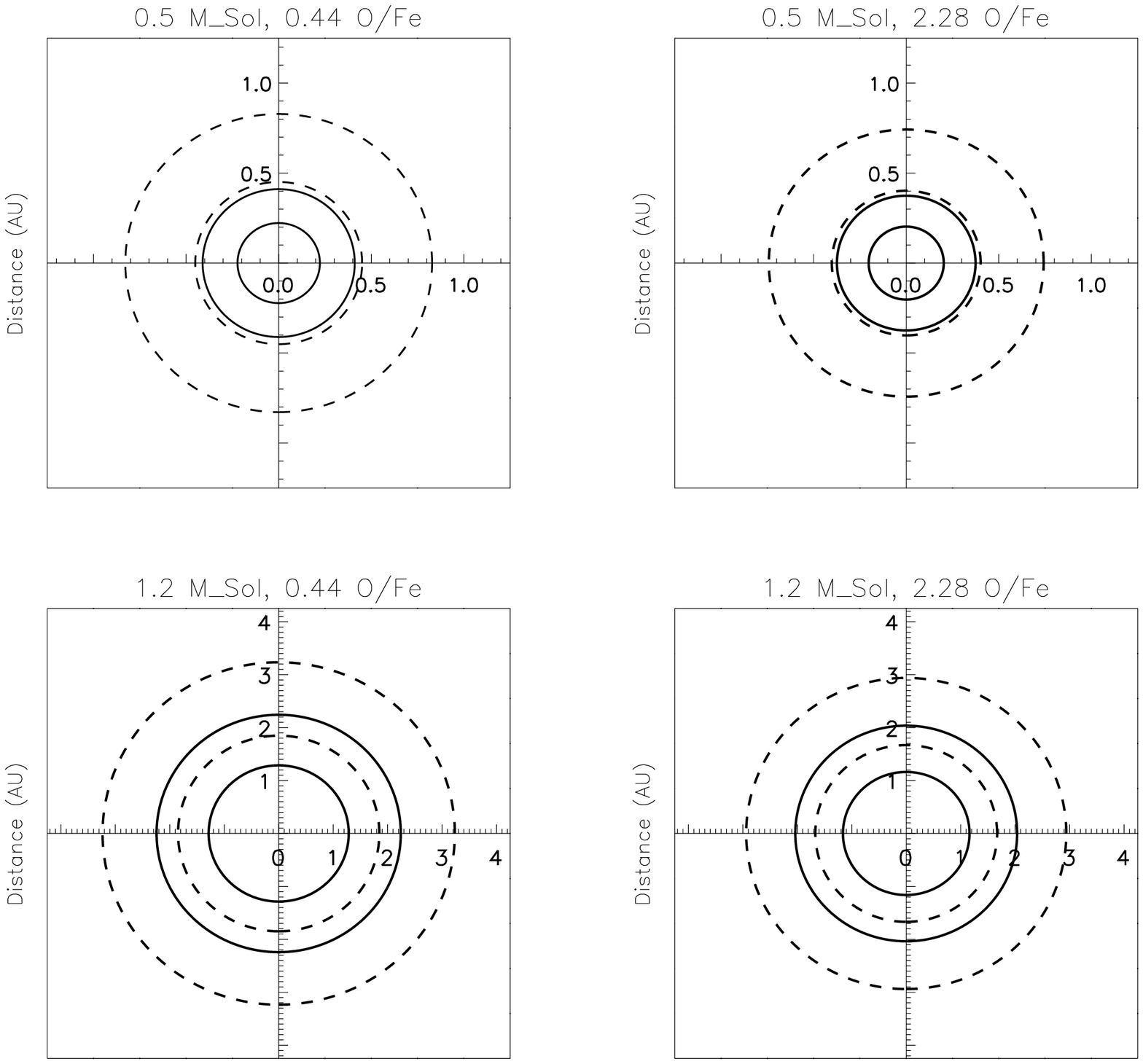}
\caption{Habitable Zone Ranges: 0.5 M\sol (top), 1.2 M\sol (bottom), for 0.44 O/Fe\sol (left), 2.28 O/Fe\sol (right), at Z\sol. HZ shown at ZAMS (solid) and TAMS (dashed). Inner boundaries represent Runaway Greenhouse and outer boundaries represent Maximum Greenhouse.\label{hzpolar2}}    
\end{figure}

Figures~\ref{hzpolar1} and \ref{hzpolar2} show the HZ from a perspective perpendicular to a hypothetical orbital plane, demonstrating how the HZ can vary between different stars. Figure~\ref{hzpolar1} shows a 0.5 M\sol star and a 1.2 M\sol star (end member masses) at the lowest and highest metallicity cases (0.1 Z\sol and 1.5 Z\sol, respectively). HZ boundaries are solid for the ZAMS and dashed for the TAMS. The high mass stars exhibit some overlap between the outer edge at the ZAMS and the inner edge at the TAMS, which might correspond to a ``Continuously Habitable Zone" (see \S 3.3). The largest overlap appears for the 0.1 Z\sol model. Given this HZ prescription, there are no orbits around the low mass stars that remain within the HZ for the entire MS. Figure~\ref{hzpolar2} similarly shows the HZ distance evolution from ZAMS to TAMS for the extreme oxygen cases (2.28 O/Fe\sol and 0.44 O/Fe\sol) at Z\sol. We see that the HZ can change substantially over the MS depending on the host star's detailed composition. Tables~\ref{tab6} and \ref{tab7} show changes in the location of the HZ radius in $\Delta$AU from ZAMS to TAMS for both the inner boundary (RGH) and the outer boundary (MaxGH), respectively.

Although there is nearly as much variation in the location of the HZ for low and high mass stars and low and high metallicity, the rate of that change has very different implications for habitability. As demonstrated in Table~\ref{tab2}, the least massive star's MS lifetime is much longer (ranging from $\sim$ 60 - 99 Gyr) than any reasonable timescale for the development of a detectable biosphere. Most orbits even near the boundaries will remain in the HZ for billions of years. Conversely, a more massive star's entire MS lifetime at low metallicity is only $\sim$ 2.8 Gyr. A planet that takes 2 Gyr or so to evolve a detectable biosphere (like Earth) would need to remain in the HZ for nearly the entire MS of the host star. Given the rate of change in the HZ position, very few orbits would qualify. 

\begin{figure}
\centering
\includegraphics[width=1\textwidth]{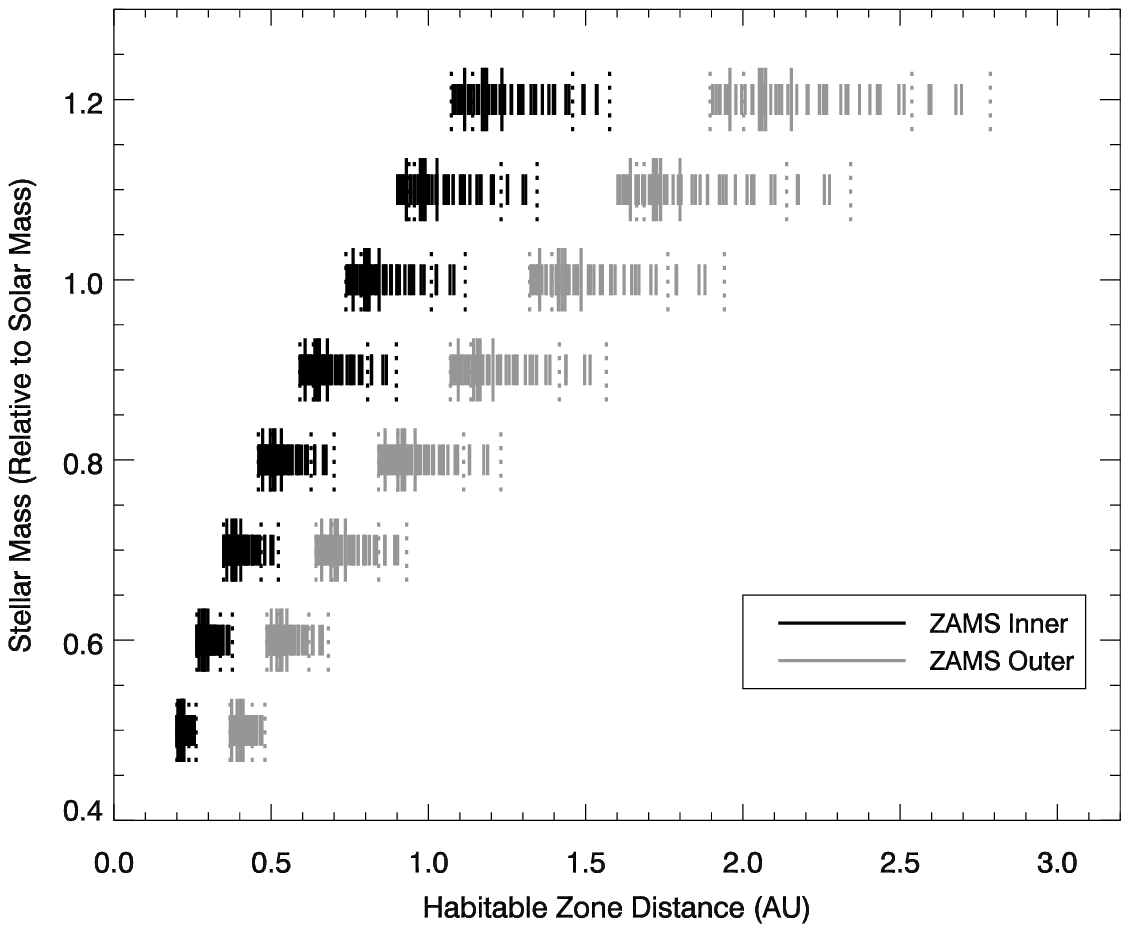}
\caption{Inner and outer HZ edges (RGH and MaxGH) for all masses and compositions at ZAMS. The elongated solid lines represent the 5 oxygen cases at Z\sol for each mass: 2 enriched oxygen models (1.48 and 2.28 O/Fe\sol), 2 depleted oxygen models (0.67 and 0.44 O/Fe\sol), and 1 model representing the standard oxygen case (solar O/Fe). The elongated dotted lines represent the end member values for the spread in oxygen abundance (0.44 and 2.28 O/Fe\sol) calculated at end member Z values (0.1 and 1.5 Z\sol). From left to right are the highest to lowest opacity models. It is clear that compositional variation has a much larger effect for the outer HZ limit, and for the higher mass stars in particular.\label{376zams}}
\end{figure}

\begin{figure}
\centering
\includegraphics[width=1\textwidth]{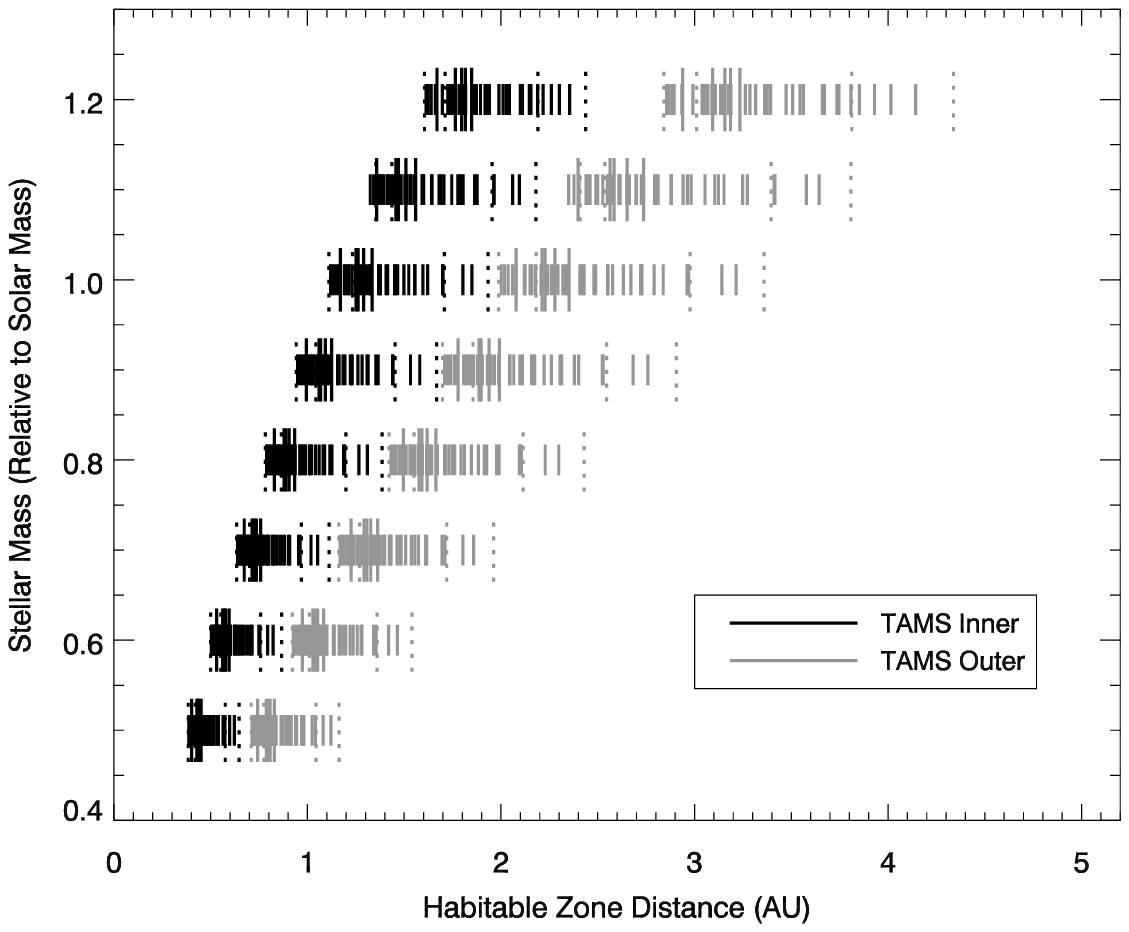}
\caption{Inner and outer HZ edges (RGH and MaxGH) for all masses and compositions at TAMS. The elongated solid lines represent the 5 oxygen cases at Z\sol for each mass: 2 enriched oxygen models (1.48 and 2.28 O/Fe\sol), 2 depleted oxygen models (0.67 and 0.44 O/Fe\sol), and 1 model representing the standard oxygen case (solar O/Fe). The elongated dotted lines represent the end member values for the spread in oxygen abundance (0.44 and 2.28 O/Fe\sol) calculated at end member Z values (0.1 and 1.5 Z\sol). From left to right are the highest to lowest opacity models. It is clear that compositional variation has a much larger effect for the outer HZ limit, and for the higher mass stars. Additionally, there exists a more exaggerated spreading trend for the lower mass stars at TAMS than occurred at ZAMS.\label{376tams}}
\end{figure}

Figure~\ref{376zams} shows the inner and outer edges of the HZ for each stellar mass for all compositions at the ZAMS, while Figure~\ref{376tams} shows the same information at the TAMS. Higher opacities result in boundaries at lower radii. It is interesting to note that with increasing mass, there seems to be a widening of the overall HZ range, as well as a larger spread in the HZ distances due to compositional variation. Figures~\ref{376zams} and~\ref{376tams} also show the spread in O variation for five values (0.44, 0.67, 1, 1.48, and 2.28 O/Fe\sol) at solar metallicity only. These models are indicated by elongated solid lines. The scaled Z cases and the varied O cases appear to be consistent in that they both show the HZ spreading out with increasing mass. The range in distance of the HZ edges for the oxygen values at Z\sol is smaller than for the entire range of metallicity. This is expected, since the total change in opacity of the stellar material is much larger for a factor of fifteen change in total metallicity than a factor of two change in oxygen abundance. Note that these figures also include the spread in oxygen calculated at $each$ metallicity value. The elongated dotted lines represent the end member values for the spread in oxygen abundance (0.44 and 2.28 O/Fe\sol) calculated at end member metallicity values (0.1 and 1.5 Z\sol). These models extend the range of HZ distance even further than do the models for oxygen abundances at Z\sol alone. The difference in HZ position as a function of composition is larger for higher mass stars because the absolute change in $L$ is larger for higher masses. The position of the outer edge changes more than that of the inner because the behavior of the Maximum Greenhouse scenario is more sensitive to the spectrum of the incoming radiation and therefore $T_{eff}$.

\begin{figure}
\centering
\includegraphics[width=0.8\textwidth]{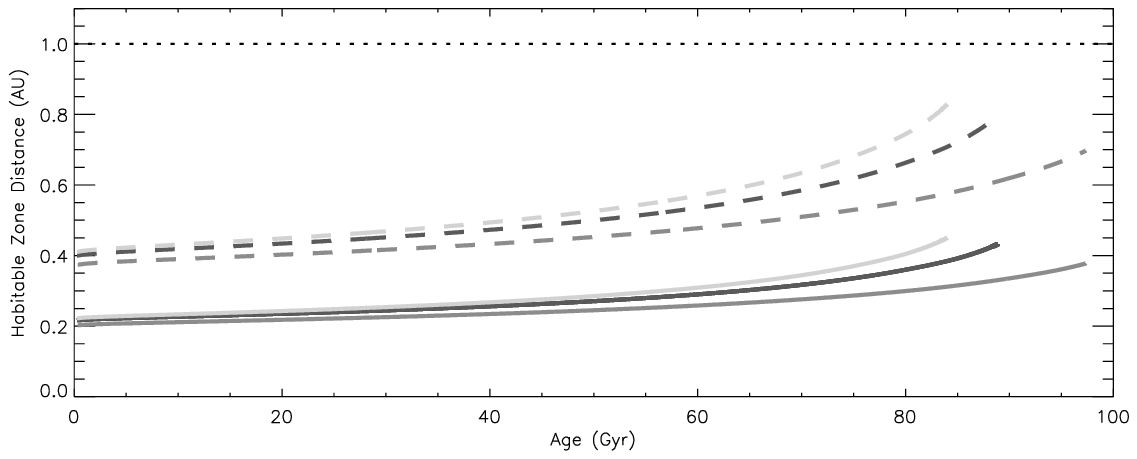}
\includegraphics[width=0.8\textwidth]{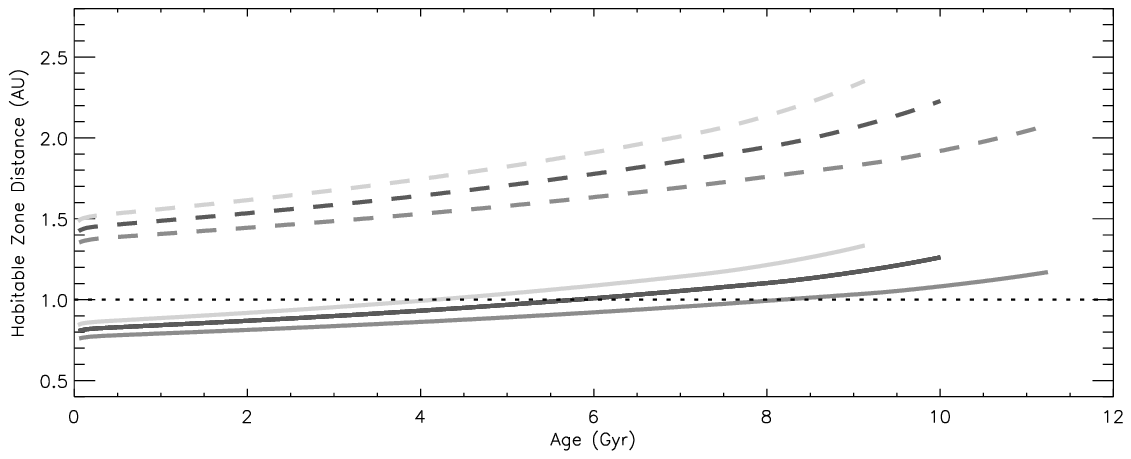}
\includegraphics[width=0.8\textwidth]{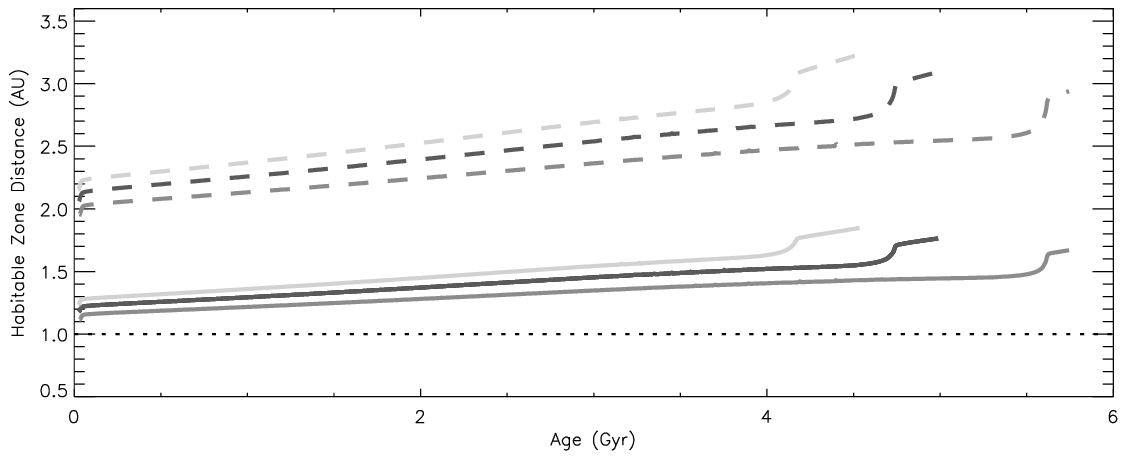}
\caption{Inner (solid) and outer (dashed) edges of the HZ for three values of O/Fe (at Z\sol) for three masses: 0.5 M\sol (top), 1 M\sol (center), and 1.2 M\sol (bottom). Each color represents a different O/Fe\sol value: black is solar O/Fe, light gray is depleted (0.44 O/Fe\sol), and dark gray is enriched (2.28 O/Fe\sol). A 1 AU orbit is indicated by the dotted line. O abundance variations within a star can significantly affect MS lifetime as well as HZ distance. The inner radius is the Runaway Greenhouse case; the outer edge is the Maximum Greenhouse case.\label{3ocase}}
\end{figure}

We produce complete evolutionary tracks for the position of the HZ as a function of time for all stellar models. With an independent age estimate for the star, as well as its mass and its composition, the position of an extrasolar planet can be compared not only to the current HZ, but also its past and future location. Assuming stellar properties are well measured, the dwell time of an observed exoplanet in the HZ can be estimated to the level of accuracy of the atmosphere models predicting HZ boundaries. Figure~\ref{3ocase} shows the location of the inner (solid) and outer (dashed) edges of the HZ as a function of time for three stellar masses (0.5, 1, and 1.2 M\sol) at Z\sol and high, standard, and low O/Fe values. The smallest radii correspond to the most enhanced oxygen model.

Again, Table~\ref{tab2} provides MS lifetimes for each stellar model of interest. The least massive (0.5 M\sol) star has the same pattern of MS lifetimes for each of the three oxygen models that we see with the higher mass stars, though the MS lifetimes are $much$ longer for the 0.5 M\sol star. The enriched oxygen model (2.28 O/Fe\sol) is estimated to live 99.3 Gyr, while the depleted oxygen model (0.44 O/Fe\sol) is estimated to live 83.5 Gyr, making the spread in lifetimes about 15.8 Gyr. Conversely, the highest mass star (1.2 M\sol) has the shortest overall MS lifetime for each of the oxygen models, with a spread of only about 1.2 Gyr between the two end-member oxygen cases. However, because the average MS lifetime of the more massive star is shorter, the percent difference in MS lifetime estimates between the end-member oxygen cases is much larger for the 1.2 M\sol star than for the 0.5 M\sol star.

Each of the stellar models represented in Table~\ref{tab2} generally have longer lifetimes as Z increases, except for the 0.5 and 0.6 M\sol stars, which show a turnover in MS lifetimes between 1 Z\sol and 1.5 Z\sol (at enriched oxygen). The percent differences between the two values is 4.85\% for the 0.5 M\sol star and 1.25\% for the 0.6 M\sol star. The reason for the turnover in lifetimes for the lowest mass stars at the highest metallicity/oxygen values is because other effects of high heavy element content become large enough to compete with enhanced opacity. In these cases, the amount of hydrogen in the core is reduced by several percent while the fraction of helium has increased. The increase in He reduces the Thompson scattering opacity of the inner, mostly ionized, region of the star while simultaneously reducing the number of free particles contributing to pressure support. Coupled with the reduced amount of fuel, the MS lifetime ends up being smaller. At even higher metallicity this trend would extend to higher masses. These factors working in tandem produce the turnover we observe, in that the MS lifetime is actually shorter for the 1.5 Z\sol case than it is for the 1 Z\sol case, both at enriched oxygen.

With more oxygen present in the host star -- that is, the host star would have a higher O/Fe ratio than present in the Sun -- the HZ will be closer in to the star because the star is less luminous and is at a lower effective temperature, and the stellar lifetime will significantly increase. Likewise, a lower oxygen abundance will produce shorter overall MS lifetimes with HZ distances that are markedly farther away from the host star. The total MS lifetime varies by about 3 Gyr between the end-member oxygen abundances. The rate of change of $L$ and $T_{eff}$ is much faster for a star with a shorter MS lifetime, and therefore the location of the HZ changes much more quickly. The habitable lifetime for a terrestrial planet varies by about 4.5 Gyr. Figure~\ref{3ocase} also demonstrates that the average lifetime for a solar mass star at solar composition (both metallicity and oxygen value) is about 10 Gyr, as expected.

\begin{figure}
\centering
\includegraphics[width=0.8\textwidth]{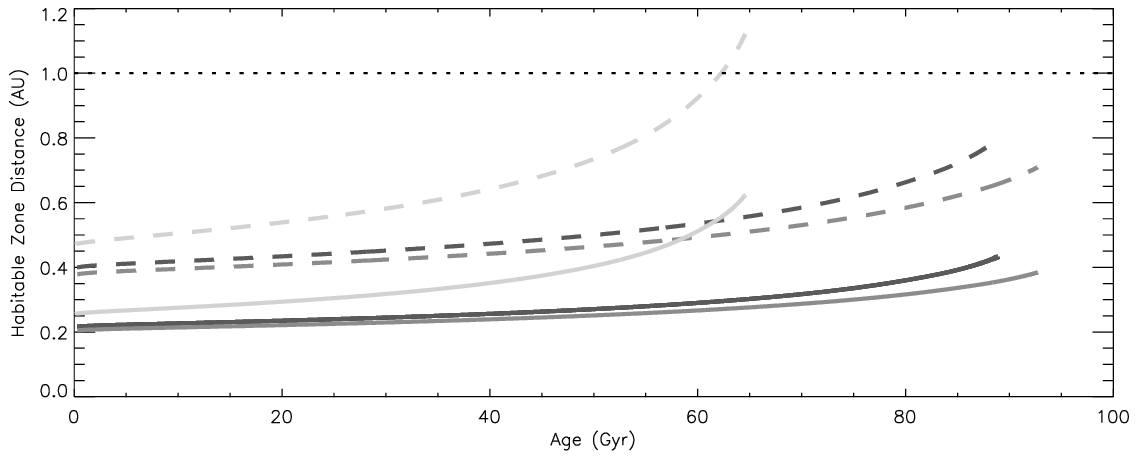}
\includegraphics[width=0.8\textwidth]{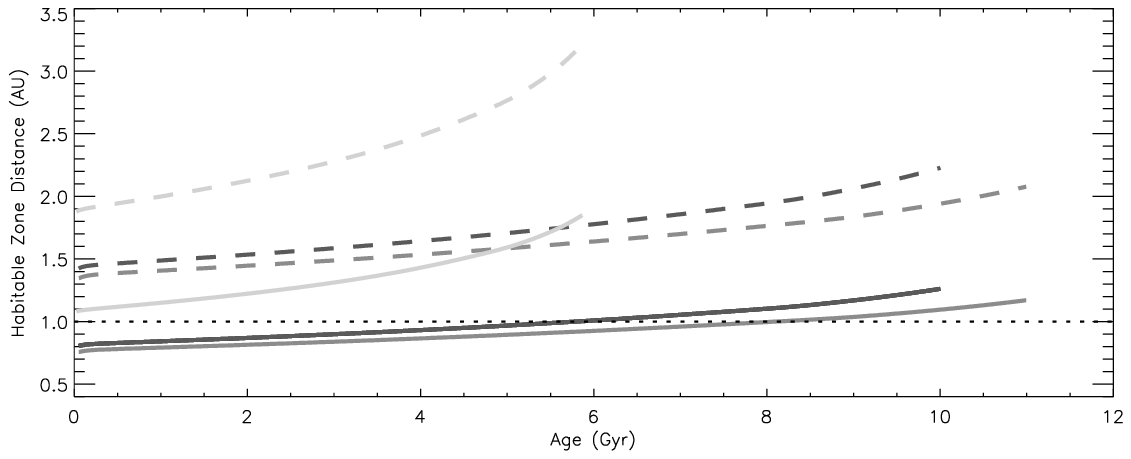}
\includegraphics[width=0.8\textwidth]{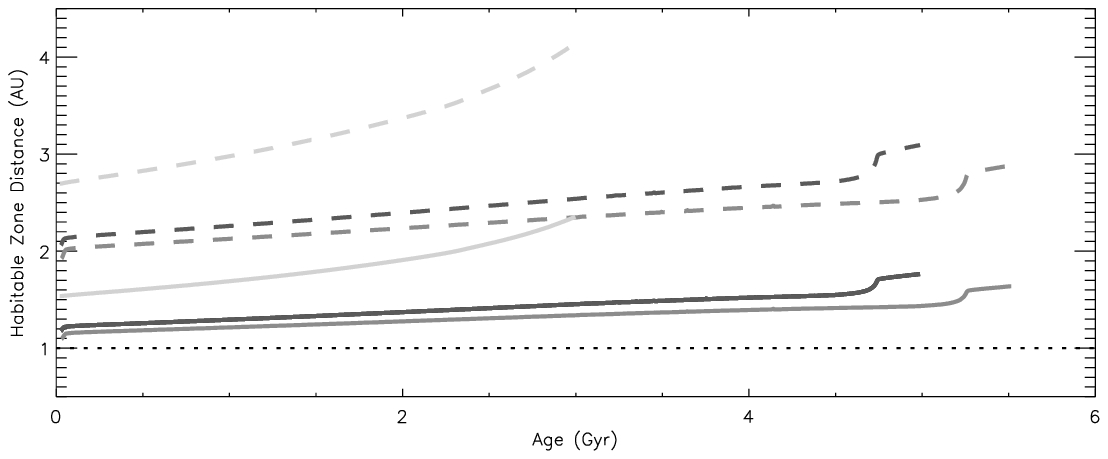}
\caption{Inner (solid) and outer (dashed) edges of the HZ at three Z values for three masses: 0.5 M\sol (top), 1 M\sol (center), and 1.2 M\sol (bottom). Each color represents a different Z value: black is Z\sol, light gray is depleted (0.1 Z\sol), and dark gray is enriched (1.5 Z\sol). A 1 AU orbit is indicated by the dotted line. Metallicity variations within a star can significantly affect MS lifetime as well as HZ distance. The largest effect is seen at 0.1 Z\sol. The inner radius is the Runaway Greenhouse case; the outer edge is the Maximum Greenhouse case.\label{3mcase}}
\end{figure}

Figure~\ref{3mcase} shows a similar trend for the end-member metallicity cases, for the same three masses (0.5, 1, and 1.2 M\sol). We find here that Earth would be interior to the HZ at all times for the lowest metallicity value. The variation is somewhat less between the high metallicity case and standard Z value, compared to the high-end oxygen and standard O value; however, they are much larger between the low metallicity case and standard value compared to standard and low oxygen. This is unsurprising, considering that a factor of 2.28 increase in O is large enough to produce an opacity increase of a similar magnitude to an overall 150\% uniform scaling of metallicity. Reducing O to 0.44 O/Fe\sol is a much smaller reduction than an overall 90\% reduction in all opacity producing elements.

A key consideration is how changing oxygen alone compares to scaling the entire metallicity of a star. For both types of compositional variation, we see the same trends in HZ distance with enhanced or depleted compositions. As overall scaled Z increases, MS lifetime increases while $L$ decreases due to greater opacity within the star, which reduces the rate at which energy can escape. Similarly, as the O abundance increases, MS lifetime increases, $L$ decreases, and the HZ will be located nearer to the host star. This is the same overall effect that metallicity exhibits, and for the same reason -- increased stellar opacity. What has not been appreciated before is that at a given [Fe/H] other individual elements can vary enough to significantly affect the stellar evolution. In fact, the higher O case at Z\sol prolongs the MS lifetime a bit more than the high scaled Z case alone.

\subsection{Continuously Habitable Zones}

\begin{figure}
\centering
\includegraphics[width=1\textwidth]{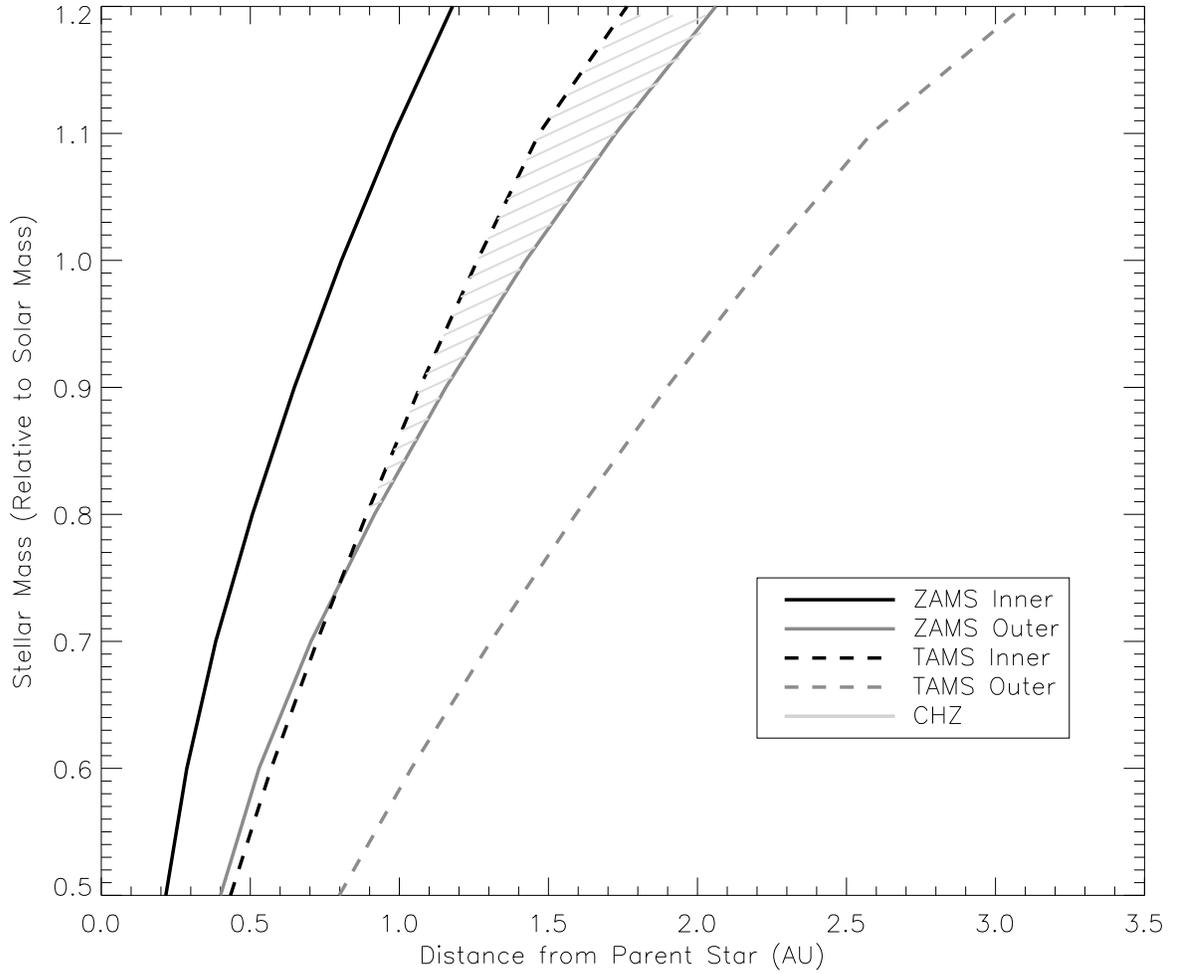}
\caption{Inner (black) and outer (dark gray) boundaries of the HZ at the ZAMS (solid) and TAMS (dashed). The light gray shaded region is the CHZ, where an orbit would remain in the HZ for the star's entire MS lifetime. This is for solar-composition stars for each mass in our range. For conservative HZ limits (RGH and MaxGH), low-mass stars have no CHZ.\label{chz}}
\end{figure}

It is useful to quantify the dwell time of a planet in the HZ as a function of its orbital radius. It is clear that the instantaneous habitability of a planet is insufficient to determine its likelihood of having life or its detectability. A planet must remain habitable for a considerable amount of time. The simplest exercise is to find a continuously habitable zone (CHZ), a range of orbital radii that remain in the HZ for the entire MS. Figure~\ref{chz} shows the CHZ for stars of all masses in our range at solar composition. The CHZ is defined by considering the boundary overlap between ZAMS and TAMS. Low mass stars have no CHZ for the conservative cases. For the optimistic case, we have overlap for the entire mass range, though it is much smaller for low mass stars than for high mass stars. This would seem to indicate that low mass stars would have a low likelihood of detectability, but this is somewhat misleading due to the long MS lifetimes of low mass stars. 

\begin{figure}
\centering
\includegraphics[width=1\textwidth]{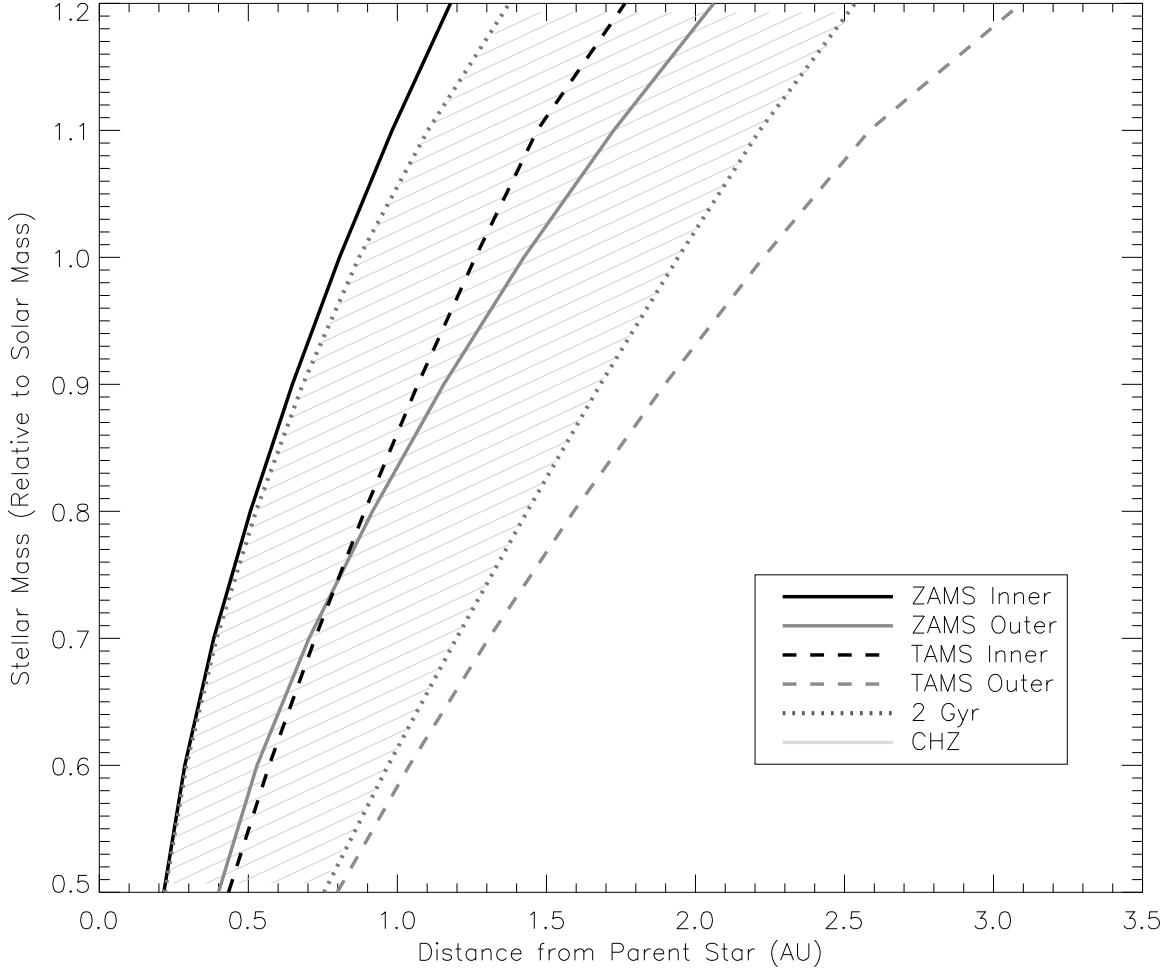}
\caption{Inner (black) and outer (dark gray) boundaries of the HZ at the ZAMS (solid) and TAMS (dashed). The inner edge 2 Gyr after the ZAMS and the outer edge 2 Gyr before the TAMS are indicated by dotted lines. The light gray shaded region is the CHZ$_2$, in which an orbiting planet would remain in the HZ for at least 2 Gyr. This is for solar-composition stars for each mass in our range. The fraction of the total habitable orbits in the CHZ$_2$ is higher for the long-lived, low mass stars.\label{chz2}}
\end{figure}

This simple version of the CHZ is of limited utility. A more useful measure would be the zone that is continuously habitable for enough time for a surface biosphere to produce a measurable and identifiable change in the planet's atmosphere. On Earth this process took  $\sim$ 2 Gyr \citep{summons99,kc03,holland06}. Here we show results for orbits that remain habitable for at least 2 Gyr (CHZ$_2$). This assumes that Earth's timescale for biological modification of the atmosphere is representative. This is not meant to suggest that other timescales are impossible, but in narrowing down a large pool of candidates there is an advantage in looking for what we know works from observing Earth's history. It is important for us to implement some kind of screening process that allows us to narrow down where any upcoming planet-finding missions should look for potentially habitable planets. It is possible of course that life may exist on non-Earth-like planets that do not fall under our 2 Gyr criteria. Figure~\ref{chz2} shows the CHZ$_2$ for standard composition. This is determined from the inner edge of the HZ at 2 Gyr after the beginning of the MS and the outer edge 2 Gyr before the TAMS, which represents the location around the star that a planet could remain habitable for at least 2 Gyr. Because of the much longer lifetimes of the low-mass stars, there is a higher proportion of the HZ included in the CHZ$_2$ than for the basic CHZ. Likewise, due to the shorter lifetimes of the more massive stars, it would be statistically less likely to find a planet orbiting this type of star that has been in the CHZ for at least 2 Gyr. We would be less confident that a planet located outside of the CHZ$_2$ would produce detectable biosignatures than one within the CHZ$_2$. With the exception of planets orbiting M-stars (because of their extremely long lifetimes), it is a useful exercise to determine the location of the CHZ$_2$. Table~\ref{tab8} shows the fraction of time a planet would spend in the CHZ$_2$ vs. time it would spend in the HZ over its entire MS lifetime. This information will help to quantify the kinds of stars we should focus on in the continued search for habitable (and possibly inhabited) exoplanets.

In our consideration of the CHZ, we must also address the issue of ``cold starts.'' The previous discussion assumes that as the HZ expands outward due to the effects of stellar evolution, any planets that were initially beyond the boundaries of the HZ could potentially become habitable as soon as the HZ reaches them; indeed, the albedos used in the planetary atmosphere models of \citet{kopp14} are relatively low, which assumes a planet could become habitable upon entering the HZ. However, it may be unlikely that a completely frozen planet (a ``hard snowball'') entering the HZ late in the host star's MS lifetime would receive enough energy in the form of stellar radiation to reverse a global glaciation, especially if the planet harbors reflective CO$_2$ clouds \citep{calkast92,kwr93}.

\begin{figure}
\centering
\includegraphics[width=0.68\textwidth]{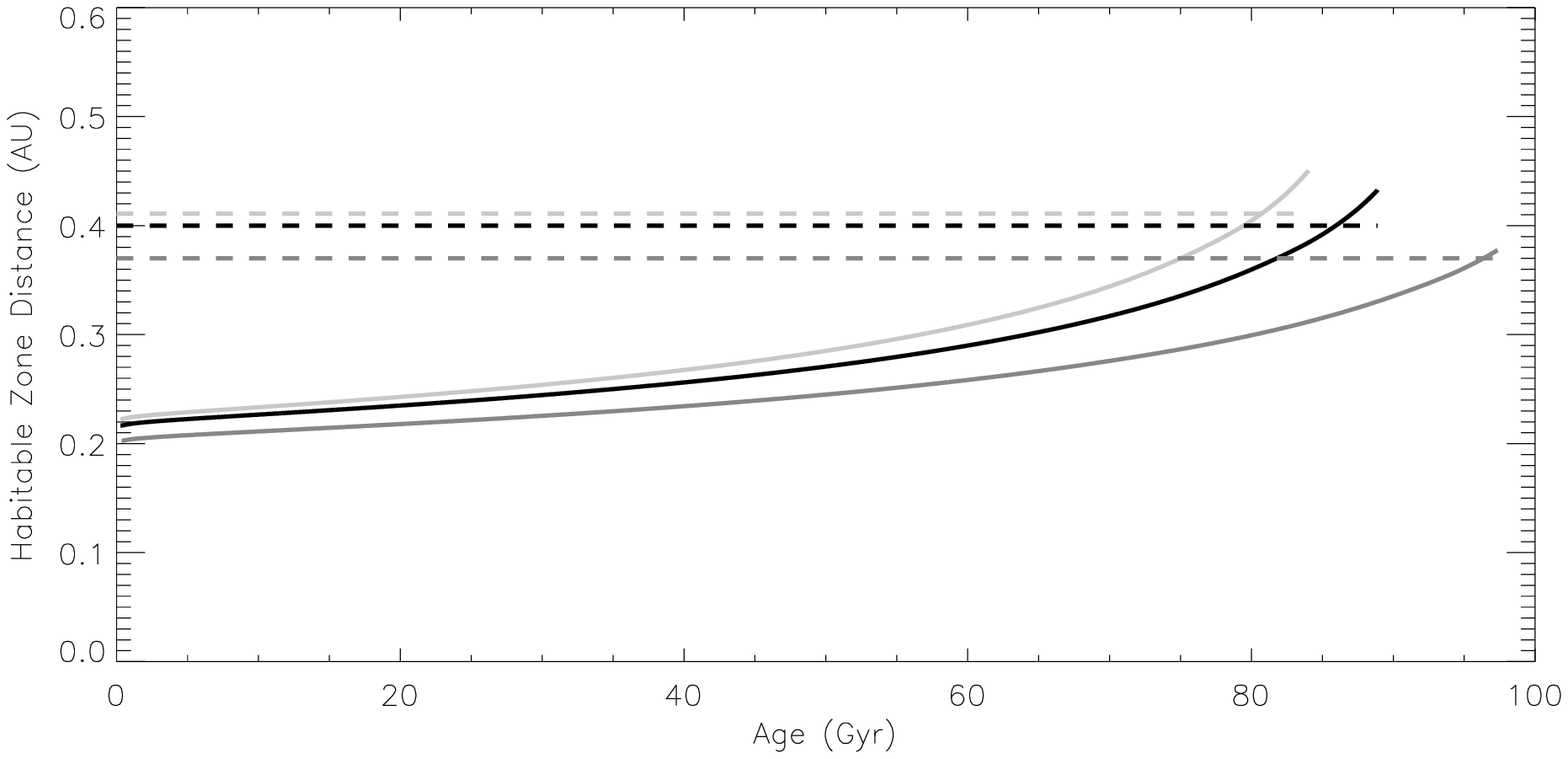}
\includegraphics[width=0.68\textwidth]{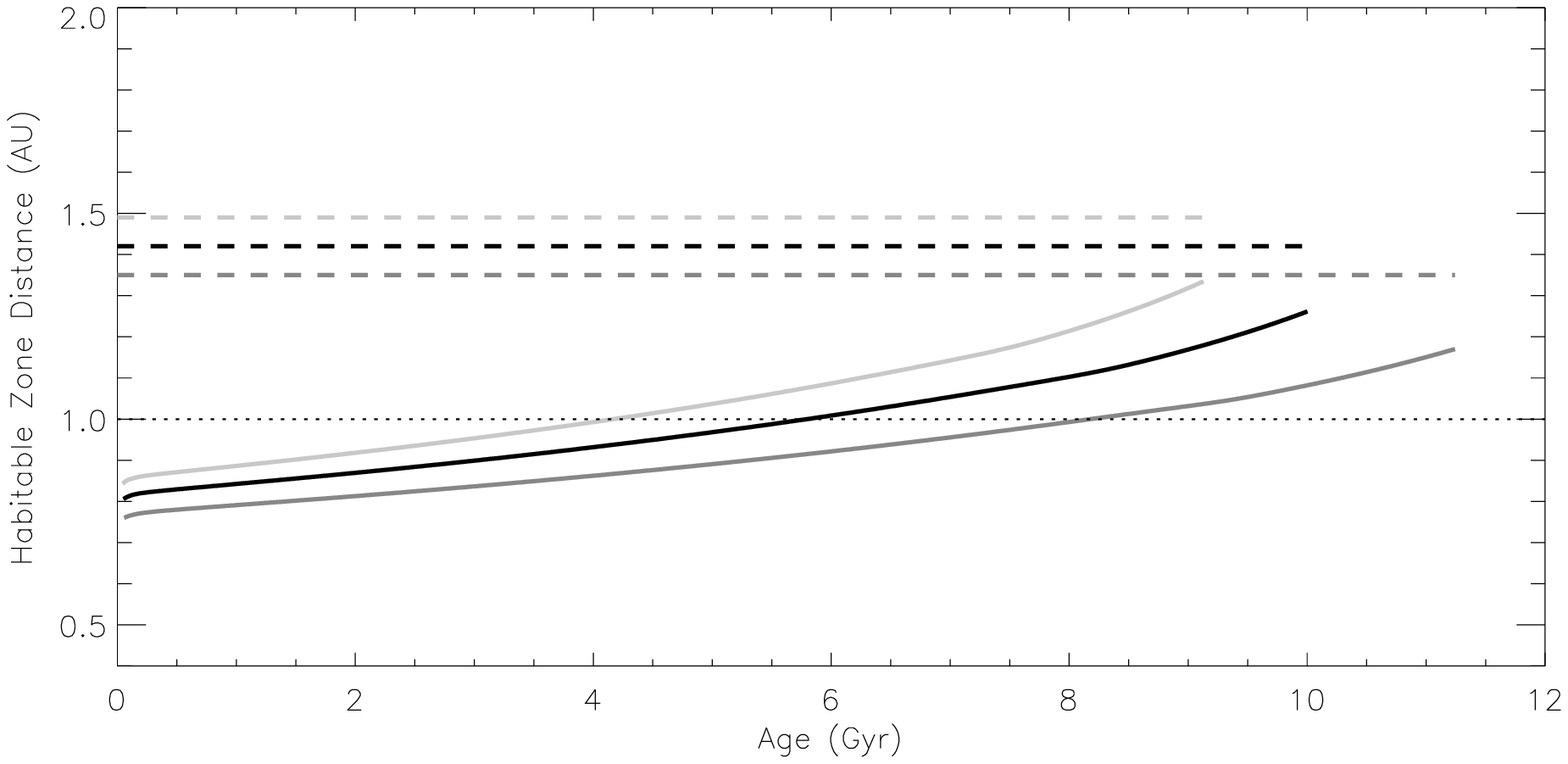}
\includegraphics[width=0.68\textwidth]{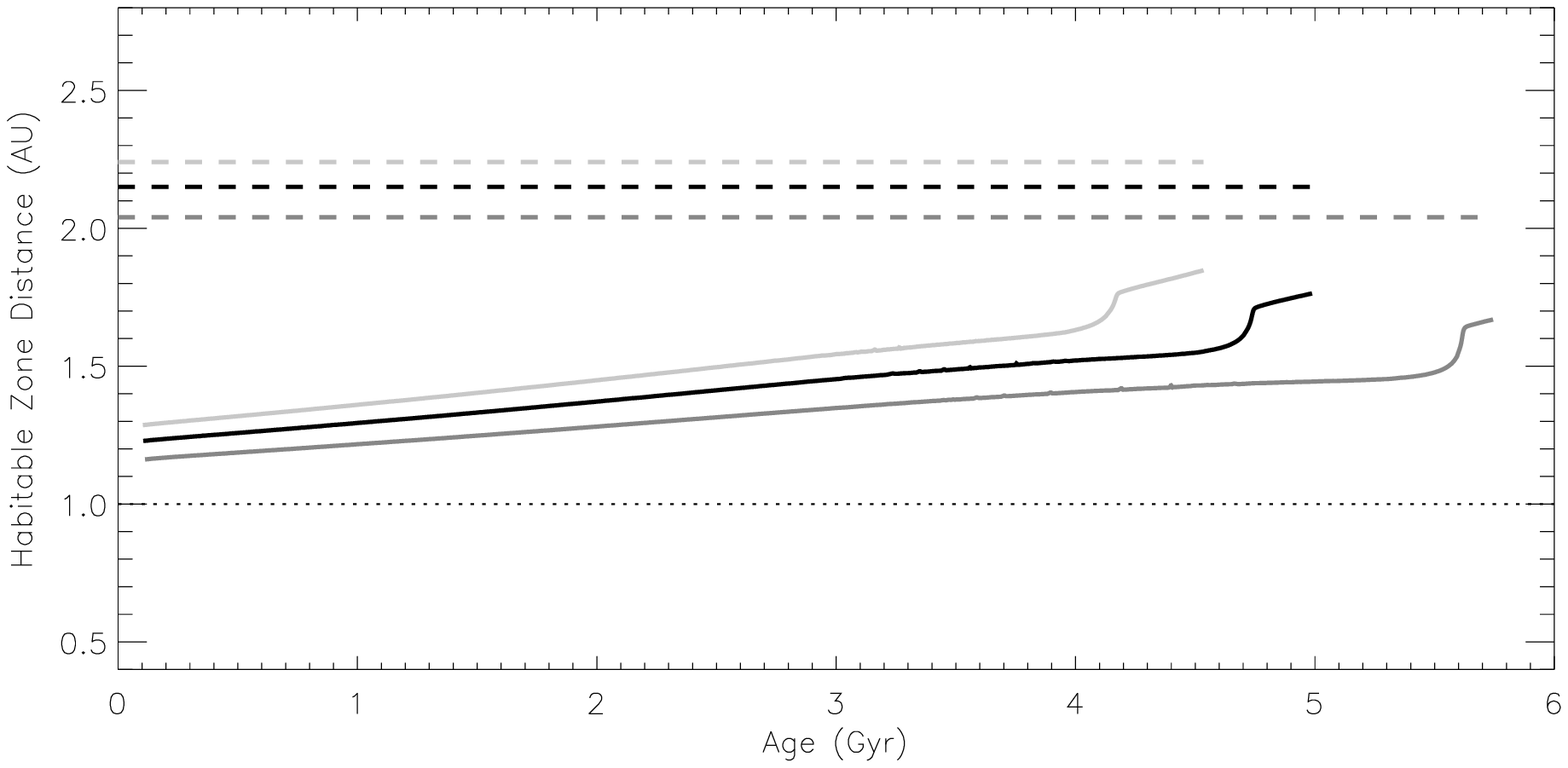}
\caption{Inner (solid) and outer (dashed) edges of the HZ with cold starts prohibited for three values of O/Fe (at Z\sol) for three masses: 0.5 M\sol (top), 1 M\sol (center), and 1.2 M\sol (bottom). Black is solar O/Fe, light gray is depleted (0.44 O/Fe\sol), and dark gray is enriched (2.28 O/Fe\sol). A 1 AU orbit is indicated by the dotted line. The inner radius is Runaway Greenhouse and the outer edge is the Maximum Greenhouse, at ZAMS value.\label{hzcold}}
\end{figure}

\begin{figure}
\centering
\includegraphics[width=1\textwidth]{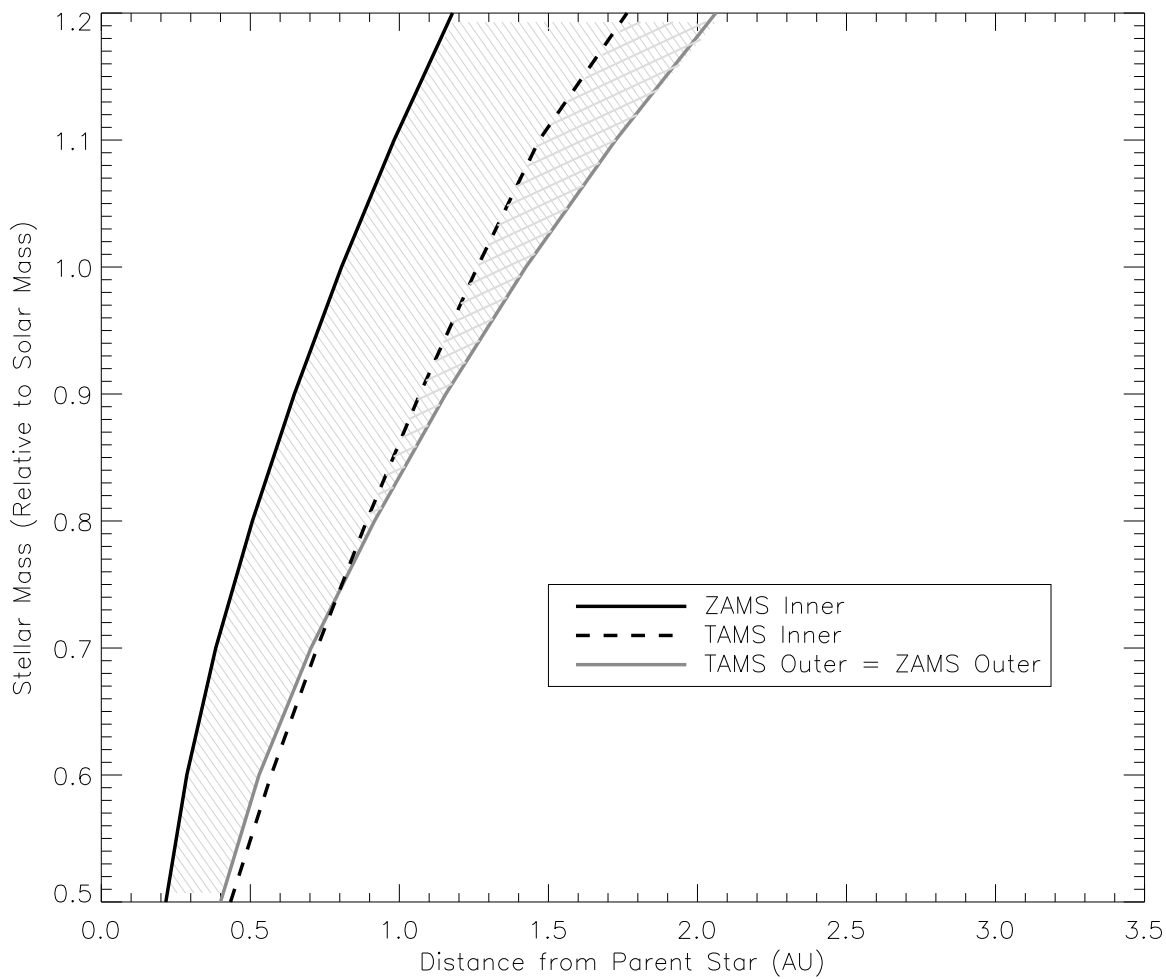}
\caption{Inner (black) and outer (dark gray) boundaries of the HZ at the ZAMS (solid) and TAMS (dashed).  This is for solar-composition stars for each mass in our range. Here we equate the TAMS outer limit with the ZAMS outer limit in order to address the problem of cold starts. The thin-lined shaded region represents the range of orbits in which a planet would be in its host star's HZ from the beginning of the MS, while the thick-lined shaded region represents the range of orbits that would be continuously habitable for the star's entire MS lifetime (as in Figure~\ref{chz}). If we attempt to understand a CHZ in this way, we see that the ideal habitable regions do not include the TAMS outer limit, where a planet could potentially start out very cold and only enter the HZ late in its lifetime.\label{chzcold}}
\end{figure}

\begin{figure}
\centering
\includegraphics[width=1\textwidth]{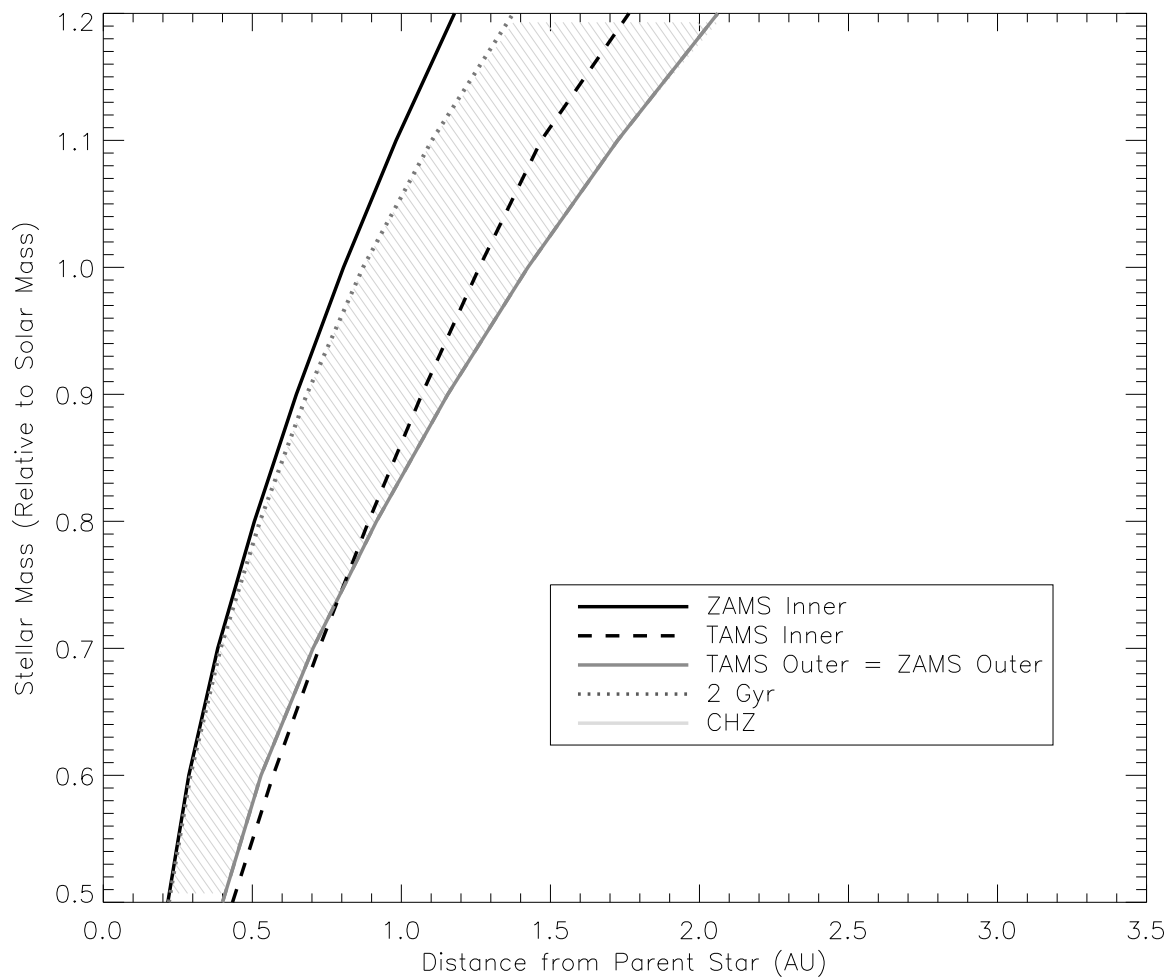}
\caption{Inner (black) and outer (dark gray) boundaries of the HZ at the ZAMS (solid) and TAMS (dashed) for solar-composition stars for each mass in our range for the case. Here we represent the case in which cold starts are prohibited. The TAMS outer limit of the HZ is equated with the ZAMS outer limit. The shaded region now represents the range of orbits in which a planet would be continuously habitable for at least 2 Gyr and would also be in its host star's HZ from the beginning of the MS.\label{chz2cold}}
\end{figure}

Figures~\ref{hzcold}, \ref{chzcold}, and \ref{chz2cold} offer an alternative scenario to that of Figures~\ref{3ocase}, \ref{chz}, and \ref{chz2}, respectively, in which we have assumed a cold start is possible. Here we treat the outer boundary of the HZ at the ZAMS as a hard limit that does not co-evolve with the host star over time, so that only planets in the HZ from the beginning of the MS will be habitable; if a planet is able to dwell within the HZ from early times, then a cold start is no longer a problem. Figure~\ref{hzcold} shows the evolution of the HZ for the same cases as Figure~\ref{3ocase}, now with cold starts prohibited. Figure~\ref{chzcold} demonstrates what the CHZ might look like if we fix the outer boundary at the ZAMS value. In this figure, the thin-lined shaded region represents the range of orbits in which a planet would simply be in its host star's HZ from the beginning of the MS, while the thick-lined shaded region represents the range of orbits that would be continuously habitable for the star's entire MS lifetime. For low mass stars there are no continuously habitable orbits, though again the long lifetimes of these stars make that less of an issue. In all cases, however, the range of orbits that could produce an inhabited planet is restricted compared to the case in which cold starts are allowed.

Similarly, Figure~\ref{chz2cold} shows the CHZ$_2$ when cold starts are prohibited, where the shaded region is now representative of the range of orbits that would be continuously habitable for at least 2 Gyr (as opposed to the entire MS lifetime), and orbits that also would be in the host star's HZ from the beginning of the MS, which would thus help us rule out any planets that would not enter the HZ until later in the star's MS lifetime. If we attempt to understand a CHZ in this way, we see that the ideal habitable regions do not include the TAMS outer limit, where a planet could potentially start out very cold and only enter the HZ late in its lifetime. Considering the time dependent evolution of the HZ is even more important in this case, since a significant fraction of planets that are observed to be within the nominal HZ now are likely to be cold start cases.

\section{CONCLUSIONS}

Given the classical definition of the Habitable Zone, which assumes a surface biosphere that supports liquid water, the properties of the host star are obviously of fundamental importance. However, ``habitable" does not automatically mean inhabited, which in turn does not equate with observable biosignatures. Many factors must be considered in calculating the likelihood for detection. The astrophysics should be treated in a more nuanced fashion, and the time evolution of the location of the HZ must also be considered.

Time evolution of the parent star is an important consideration, because the timescale for life to develop to a point where it alters the planetary atmosphere sufficiently for biogenic non-equilibrium species to be detectable may be quite long. On Earth, this process took $\sim$ 2 Gyr. Some attention has been paid to this aspect, but it is overlooked surprisingly often when announcements are made of planets in the HZ of the associated host star. Additionally, the concept of a CHZ is too simple, since it does not take into account the variable lifetimes of stars. There is no CHZ for a solar composition 0.5 M\sol star (using the conservative HZ boundary estimates), but a set of orbits need only be habitable for a small fraction of the total stellar lifetime (which is on the order of 10$^{11}$ years) to meet the 2 Gyr criterion. We propose a 2 Gyr continuously habitable zone as an aid to estimating the likelihood for detection, which is a necessary, but not completely sufficient, condition. Because of the shape of the luminosity vs. time curve for a star, a third to a half of habitable orbits only become so in the second half of the star's life on the MS. For example, one can certainly imagine planets in these orbits entering the HZ when they are geologically dead, which may be unfavorable to life (see Table~\ref{tab8}).

In turn, the evolution of a star depends on its elemental composition. This is a basic fact of stellar astrophysics and has been considered in HZ evolution in terms of total metallicity by some groups (e.g. \citet{dl13}). However, it is important to distinguish between the metallicity of a star as measured by [Fe/H] and the actual abundances of individual elements. Metallicity is often used interchangeably with [Fe/H], which expresses the amount of iron alone relative to the Sun's elemental abundance. The standard approach to stellar modeling is typically to scale the metallicity assuming all elements scale in the same proportions found in the Sun, though observations show that this is very often not the case in reality. The HZ distance can be substantially affected even when only abundance {\it ratios} are changed. Evaluating habitability by modeling stellar and HZ evolution requires models that span a range of variation in abundance ratios, as well as total metallicity. For the same reason, characterizing a system requires measurements of multiple elemental abundances, not just [Fe/H].

Here we present the results of changing O/Fe at levels similar to those seen in nearby stars as well as the total metallicity. Both changes influence the evolution by changing the stellar opacity and therefore have similar effects.  For low metallicity or low O/Fe at a given [Fe/H], MS lifetimes are shorter, and the total luminosity change over the MS is larger. This results in a high $dL/dt$ and $dT_{eff}/dt$, and a correspondingly rapid change in the location of the HZ. What is not widely appreciated is that changes in abundance ratios can have very large effects. For example, variation of O/Fe values from a base solar composition by an amount observed in some planet host candidates can change the stellar lifetime more than increasing the metallicity by 50\%. An increase in O increases the total heavy element abundance, but there are two important differences. First, the opacity will be different for different mixes of elements with the same integrated heavy element abundances. Second, as long as model comparisons with individual stars are made based only on the measured [Fe/H], which is the common practice, very large errors may be present.

Since many targets of radial velocity planet searches have high quality spectra that can be used to determine abundances, it should not be difficult to compare to stellar models with more accurate compositions as long as such models exist. We have made the library of stellar evolution models and predicted HZs upon which this paper is based available to the community at http://bahamut.sese.asu.edu/$\sim$payoung/AST\_522/Evolutionary\_Tracks\_Database.html. A routine for interpolation between tracks for any specified model contained within the grid boundaries will also be provided. The library will be extended in the future for additional elements and masses, as well as later evolutionary stages for sufficiently short-lived stars.

{\bf Acknowledgements} A. Truitt and P. Young gratefully acknowledge support from NASAÕs Follow the Elements team under cooperative agreement NNH08ZDA002C. The authors would like to thank Mike Pagano for information on stellar compositions and Jerry Dietrich and Suzanne W. Dietrich for advice on setting up the stellar track database.

\clearpage

\begin{deluxetable}{lc}
\tablewidth{0pt}
\tablecaption{Abundance adjustments from solar composition to average sample composition.\label{tab1}}
\tablehead{
  \colhead{Element}
& \colhead{[X/H]$_{standard}$\tablenotemark{a}}
}

\startdata
C  &  0.173 \\
O  &  0.051\\
Na  &  0.068 \\
Mg  &  -0.029\\
Al  &  -0.124 \\
Si  &  0.078 \\
Ca  &  0.027\\
Ti  &  0.072\\

\enddata
\tablenotetext{a}{Standard composition values relative to solar value \citep{lod10}}
\end{deluxetable}

\begin{deluxetable}{lcccccccc}
\tabletypesize{\scriptsize}
\tablecaption{MS lifetimes (Gyr) at standard and end member oxygen values, at solar and end member metallicity values, for all masses. \label{tab2}}
\tablehead{
  \colhead{Composition}
& \colhead{0.5 M\sol}
& \colhead{0.6 M\sol}
& \colhead{0.7 M\sol}
& \colhead{0.8 M\sol}
& \colhead{0.9 M\sol}
& \colhead{1.0 M\sol}
& \colhead{1.1 M\sol}
& \colhead{1.2 M\sol}
}
\startdata
0.1 $Z\sol$, 0.44 O/Fe\sol  &  60.4  &  33.6  &  19.8  &  12.3  &  8.0  &  5.5  &  3.9  &  2.8 \\
0.1 $Z\sol$, O/Fe\sol  &  64.2  &  35.9  &  21.2  &  13.1  &  8.6  &  5.8  &  4.1  &  3.0 \\
0.1 $Z\sol$, 2.28 O/Fe\sol  &  76.9  &  43.3  &  25.6  &  15.7  &  10.3  &  6.9  &  4.8  &  3.5 \\

$Z\sol$, 0.44 O/Fe\sol  & 83.5  &  52.8  &  32.5  &  20.6  &  13.5  &  9.1  &  6.2  &  4.5 \\
$Z\sol$, O/Fe\sol  &  88.4  &  57.3  &  35.8  &  22.7  &  14.8  &  9.9  &  6.7  &  4.9 \\
$Z\sol$, 2.28 O/Fe\sol  &  99.3  &  64.4  &  41.2  &  26.1  &  16.9  &  11.1  &  7.5  &  5.7 \\

1.5 $Z\sol$, 0.44 O/Fe\sol  &  87.6  &  57.6  &  36.5  &  23.3  &  15.3  &  10.3  &  7.0  &  5.1 \\
1.5 $Z\sol$, O/Fe\sol  &  93.6  &  60.8  &  39.0  &  24.9  &  16.3  &  10.9  &  7.4  &  5.5 \\
1.5 $Z\sol$, 2.28 O/Fe\sol  &  94.6  &  63.6  &  41.2  &  26.2  &  17.0  &  11.1  &  7.9  &  5.8 \\
\enddata
\end{deluxetable}

\clearpage

\begin{deluxetable}{lllllllll}
\tabletypesize{\scriptsize}
\tablecaption{$\Delta$($L$/$L_{ZAMS}$) for each mass and end-member composition.\label{tab3}}
\tablehead{
  \colhead{Composition}
& \colhead{0.5 M\sol}
& \colhead{0.6 M\sol}
& \colhead{0.7 M\sol}
& \colhead{0.8 M\sol}
& \colhead{0.9 M\sol}
& \colhead{1.0 M\sol}
& \colhead{1.1 M\sol}
& \colhead{1.2 M\sol}
}
\startdata
0.1 $Z\sol$, 0.44 O/Fe\sol  &  5.64  &  4.62  &  3.70  &  2.99  &  2.48  &  2.04  &  1.69  &  1.43 \\
0.1 $Z\sol$, O/Fe\sol  &  5.41  &  4.41  &  3.51  &  2.84  &  2.35  &  1.94  &  1.60  &  1.37 \\
0.1 $Z\sol$, 2.28 O/Fe\sol  &  5.27  &  4.40  &  3.50  &  2.73  &  2.28  &  1.87  &  1.52  &  1.27 \\

$Z\sol$, 0.44 O/Fe\sol  & 3.27  &  3.21  &  2.73  &  2.19  &  1.79  &  1.50  &  1.13  &  1.00 \\
$Z\sol$, O/Fe\sol  &  3.17  &  3.16  &  2.76  &  2.21  &  1.78  &  1.46  &  1.09  &  1.01 \\
$Z\sol$, 2.28 O/Fe\sol  &  3.07  &  3.00  &  2.71  &  2.19  &  1.74  &  1.39  &  1.01  &  1.02 \\

1.5 $Z\sol$, 0.44 O/Fe\sol  &  2.96  &  2.98  &  2.64  &  2.13  &  1.75  &  1.47  &  1.05  &  0.94 \\
1.5 $Z\sol$, O/Fe\sol  &  2.89  &  2.84  &  2.57  &  2.11  &  1.71  &  1.41  &  1.01  &  0.94 \\
1.5 $Z\sol$, 2.28 O/Fe\sol  &  2.81  &  2.75  &  2.45  &  1.99  &  1.58  &  1.28  &  1.05  &  0.91 \\
\enddata
\end{deluxetable}

\begin{deluxetable}{lcccccccc}
\tabletypesize{\scriptsize}
\tablecaption{$\Delta$$T$ (K) for each mass and end-member composition.\label{tab4}}
\tablehead{
  \colhead{Composition}
& \colhead{0.5 M\sol}
& \colhead{0.6 M\sol}
& \colhead{0.7 M\sol}
& \colhead{0.8 M\sol}
& \colhead{0.9 M\sol}
& \colhead{1.0 M\sol}
& \colhead{1.1 M\sol}
& \colhead{1.2 M\sol}
}
\startdata
0.1 $Z\sol$, 0.44 O/Fe\sol  &  719  &  555  &  315  &  136  &  58  &  101  &  217  &  168 \\
0.1 $Z\sol$, O/Fe\sol  &  718  &  592  &  344  &  149  &  53  &  29  &  125  &  114 \\
0.1 $Z\sol$, 2.28 O/Fe\sol  &  652  &  665  &  425  &  171  &  82  &  22  &  -5  &  41 \\

$Z\sol$, 0.44 O/Fe\sol  &  320  &  502  &  495  &  303  &  118  &  -15  &  -137  &  -243 \\
$Z\sol$, O/Fe\sol  &  303  &  463  &  511  &  346  &  170  &  30  &  -100  &  -215 \\
$Z\sol$, 2.28 O/Fe\sol  &  256  &  349  &  439  &  346  &  202  &  78  &  -57  &  -193 \\ 

1.5 $Z\sol$, 0.44 O/Fe\sol  &  277  &  417  &  475  &  335  &  170  &  44  &  -128  &  -244 \\
1.5 $Z\sol$, O/Fe\sol  &  240  &  345  &  430  &  330  &  193  &  76  &  -74  &  -198 \\
1.5 $Z\sol$, 2.28 O/Fe\sol  &  248  &  296  &  344  &  269  &  144  &  50  &  -126  &  -237 \\
\enddata
\end{deluxetable}

\begin{deluxetable}{lcccccccc}
\tabletypesize{\scriptsize}
\tablecaption{Fraction (\%) of radii which are in the HZ after the midpoint of the MS.\label{tab5}}
\tablehead{
  \colhead{Composition}
& \colhead{0.5 M\sol}
& \colhead{0.6 M\sol}
& \colhead{0.7 M\sol}
& \colhead{0.8 M\sol}
& \colhead{0.9 M\sol}
& \colhead{1.0 M\sol}
& \colhead{1.1 M\sol}
& \colhead{1.2 M\sol}
}
\startdata
0.1 $Z\sol$, 0.44 O/Fe\sol  &  30.6  &  33.1  &  35.4  &  37.9  &  40.8  &  44.3  &  48.2  &  52.0 \\
0.1 $Z\sol$, O/Fe\sol  &  31.1  &  33.8  &  36.4  &  39.0  &  41.8  &  45.1  &  49.0  &  52.5 \\
0.1 $Z\sol$, 2.28 O/Fe\sol  &  31.4  &  34.5  &  37.1  &  40.0  &  42.8  &  46.1  &  49.7  &  53.5 \\

$Z\sol$, 0.44 O/Fe\sol  & 37.7  &  39.1  &  42.1  &  45.1  &  48.1  &  51.1  &  55.0  &  56.1 \\
$Z\sol$, O/Fe\sol  &  38.1  &  39.2  &  42.0  &  45.3  &  48.5  &  51.9  &  56.2  &  56.4 \\
$Z\sol$, 2.28 O/Fe\sol  &  38.4  &  39.6  &  42.2  &  45.8  &  49.5  &  53.3  &  58.2  &  56.6 \\

1.5 $Z\sol$, 0.44 O/Fe\sol  &  39.4  &  40.2  &  42.9  &  46.1  &  48.9  &  52.0  &  56.6  &  57.3 \\
1.5 $Z\sol$, O/Fe\sol  &  39.5  &  40.7  &  43.1  &  46.4  &  49.8  &  53.1  &  57.7  &  57.8 \\
1.5 $Z\sol$, 2.28 O/Fe\sol  &  40.0  &  41.1  &  43.8  &  47.5  &  51.5  &  55.4  &  56.5  &  58.0 \\
\enddata
\end{deluxetable}

\clearpage

\begin{deluxetable}{lcccccccc}
\tabletypesize{\scriptsize}
\tablecaption{$\Delta$AU for each mass and end-member composition at inner HZ limit (RGH).\label{tab6}}
\tablehead{
  \colhead{Composition}
& \colhead{0.5 M\sol}
& \colhead{0.6 M\sol}
& \colhead{0.7 M\sol}
& \colhead{0.8 M\sol}
& \colhead{0.9 M\sol}
& \colhead{1.0 M\sol}
& \colhead{1.1 M\sol}
& \colhead{1.2 M\sol}
}
\startdata
0.1 $Z\sol$, 0.44 O/Fe\sol  &  0.39  &  0.49  &  0.59  &  0.69  &  0.77  &  0.82  &  0.83  &  0.86 \\
0.1 $Z\sol$, O/Fe\sol  &  0.37  &  0.46  &  0.55  &  0.63  &  0.71  &  0.77  & 0.79  & 0.82 \\
0.1 $Z\sol$, 2.28 O/Fe\sol  &  0.34  & 0.42  & 0.50  &  0.57  &  0.65  &  0.70  &  0.72  &  0.73 \\

$Z\sol$, 0.44 O/Fe\sol  &  0.23  &  0.30  &  0.35  &  0.40  &  0.45  &  0.49  &  0.50  &  0.56\\
$Z\sol$, O/Fe\sol & 0.22  & 0.28  & 0.34  & 0.38  & 0.42  & 0.46  & 0.46  & 0.54 \\
$Z\sol$, 2.28 O/Fe\sol &  0.20  &  0.26  &  0.31  &  0.36  &  0.39  &  0.409  &  0.403  &  0.51 \\ 

1.5 $Z\sol$, 0.44 O/Fe\sol  &  0.21  &  0.27  &  0.32  &  0.37  &  0.41  &  0.45  &  0.44  &  0.50 \\
1.5 $Z\sol$, O/Fe\sol  &  0.19  &  0.25  &  0.31  &  0.35  &  0.38  &  0.41  &  0.41  &  0.48 \\
1.5 $Z\sol$, 2.28 O/Fe\sol  &  0.18  &  0.24  &  0.29  &  0.32  &  0.35  &  0.37  &  0.42  &  0.46 \\
\enddata
\end{deluxetable}

\begin{deluxetable}{lcccccccc}
\tabletypesize{\scriptsize}
\tablecaption{$\Delta$AU for each mass and end-member composition at outer HZ limit (MaxGH).\label{tab7}}
\tablehead{
  \colhead{Composition}
& \colhead{0.5 M\sol}
& \colhead{0.6 M\sol}
& \colhead{0.7 M\sol}
& \colhead{0.8 M\sol}
& \colhead{0.9 M\sol}
& \colhead{1.0 M\sol}
& \colhead{1.1 M\sol}
& \colhead{1.2 M\sol}
}
\startdata
0.1 $Z\sol$, 0.44 O/Fe\sol  &  0.68 &  0.86  &  1.03  &  1.20  &  1.34  &  1.42  &  1.47  &  1.55 \\
0.1 $Z\sol$, O/Fe\sol  &  0.65  &  0.80  &  0.96  &  1.11  &  1.24  &  1.34  &  1.37 &  1.45 \\
0.1 $Z\sol$, 2.28 O/Fe\sol  &  0.61  &  0.74  &  0.88  &  1.00  &  1.13  &  1.22  &  1.26  &  1.28 \\

$Z\sol$, 0.44 O/Fe\sol  &  0.42  &  0.53  &  0.63  &  0.71  &  0.79  &  0.87  &  0.88  &  0.99 \\
$Z\sol$, O/Fe\sol  &  0.40  &  0.51  &  0.60  &  0.68  &  0.74  &  0.80  &  0.81  &  0.95 \\
$Z\sol$, 2.28 O/Fe\sol  &  0.37  &  0.47  &  0.57  &  0.63  &  0.68  &  0.72  &  0.71  &  0.90 \\ 

1.5 $Z\sol$, 0.44 O/Fe\sol  &  0.38  &  0.49  &  0.58  &  0.65  &  0.72  &  0.79  &  0.78  &  0.89 \\
1.5 $Z\sol$, O/Fe\sol  &  0.36  &  0.46  &  0.55  &  0.62  &  0.68  &  0.73  &  0.72  &  0.85 \\
1.5 $Z\sol$, 2.28 O/Fe\sol  &  0.34  &  0.44  &  0.52  &  0.58  &  0.63  &  0.67  &  0.75  &  0.83 \\
\enddata
\end{deluxetable}

\begin{deluxetable}{lcccccccc}
\tabletypesize{\scriptsize}
\tablecaption{Fraction (\%) of time spent in CHZ${_2}$ vs. the entire MS.\label{tab8}}
\tablehead{
  \colhead{Composition}
& \colhead{0.5 M\sol}
& \colhead{0.6 M\sol}
& \colhead{0.7 M\sol}
& \colhead{0.8 M\sol}
& \colhead{0.9 M\sol}
& \colhead{1.0 M\sol}
& \colhead{1.1 M\sol}
& \colhead{1.2 M\sol}
}
\startdata
0.1 $Z\sol$, 0.44 O/Fe\sol  &  85.8  &  81.0  &  75.2  &  69.0  &  61.6  &  54.3  &  0.0\tablenotemark{a}  &  0.0 \\
0.1 $Z\sol$, O/Fe\sol  &  86.8  &  82.6  &  77.5  &  71.5  &  64.6  &  57.3  &  50.0  &  0.0 \\
0.1 $Z\sol$, 2.28 O/Fe\sol  &  88.7  &  85.1  &  81.0  &  75.9  &  69.7  &  63.0  &  55.8  &  0.0 \\

$Z\sol$, 0.44 O/Fe\sol  & 91.5  &  89.2  &  86.2  &  82.6  &  78.2  &  73.0  &  67.4  &  59.5 \\
$Z\sol$, O/Fe\sol  &  91.8  &  89.9  &  87.2  &  83.9  &  79.9  &  75.3  &  70.7  &  62.2 \\
$Z\sol$, 2.28 O/Fe\sol  &  92.7  &  90.9  &  88.8  &  86.0  &  82.5  &  78.2  &  74.4  &  65.2 \\

1.5 $Z\sol$, 0.44 O/Fe\sol  &  92.0  &  90.0  &  87.6  &  84.5  &  80.3  &  75.5  &  71.2  &  63.4 \\
1.5 $Z\sol$, O/Fe\sol  &  92.7  &  90.8  &  88.5  &  85.6  &  81.8  &  77.0  &  73.1  &  65.1 \\
1.5 $Z\sol$, 2.28 O/Fe\sol  &  92.8  &  91.3  &  89.4  &  86.8  &  83.3  &  78.7  &  70.6  &  65.7 \\
\enddata
\tablenotetext{a}{No orbits are continually habitable for 2 Gyr as a result of the short MS lifetime.}
\end{deluxetable}

\clearpage

\clearpage

\clearpage

\clearpage

\clearpage

\clearpage

\clearpage

\clearpage

\clearpage

\clearpage

\clearpage

\clearpage

\clearpage

\clearpage

\clearpage

\clearpage

{}

\end{document}